\documentclass[11pt]{article}

\usepackage{cancel}
\usepackage{cite}
\usepackage{slashed}
\usepackage{amsmath,bbm}
\usepackage{amssymb}
\usepackage{graphicx}
\usepackage[latin1]{inputenc}
\usepackage{mathrsfs}
\usepackage[dvips]{color}
\usepackage[footnotesize]{caption}

\newcommand{\sw}{s^2_W}
\newcommand{\eps}{\varepsilon}
\newcommand{\bea}{\begin{eqnarray}}
\newcommand{\eea}{\end{eqnarray}}
\newcommand{\be}{\begin{equation}}
\newcommand{\ee}{\end{equation}}
\newcommand{\ba}{\begin{array}}
\newcommand{\ea}{\end{array}}

\def\gsim{\mathrel{\rlap{\lower4pt\hbox{\hskip1pt$\sim$}}
    \raise1pt\hbox{$>$}}}

\begin{document}

\begin{titlepage}

\vspace*{-15mm}
\begin{flushright}
MPP-2015-24\\

\end{flushright}
\vspace*{0.7cm}

\begin{center}
{
\bf\Large
Testing sterile neutrino extensions of the  \\[1mm] Standard Model at future lepton colliders
}
\\[8mm]
Stefan~Antusch$^{\star\dagger}$
\footnote{E-mail: \texttt{stefan.antusch@unibas.ch}}
and 
Oliver~Fischer$^{\star}$
\footnote{E-mail: \texttt{oliver.fischer@unibas.ch}},
\\[1mm]
\end{center}
\vspace*{0.50cm}
\centerline{$^{\star}$ \it
 Department of Physics, University of Basel,}
\centerline{\it
Klingelbergstr.~82, CH-4056 Basel, Switzerland}
\vspace*{0.2cm}
\centerline{$^{\dagger}$ \it
Max-Planck-Institut f\"ur Physik (Werner-Heisenberg-Institut),}
\centerline{\it
F\"ohringer Ring 6, D-80805 M\"unchen, Germany}
\vspace*{1.20cm}
\begin{abstract}
\noindent
Extending the Standard Model (SM) with sterile (``right-handed'') neutrinos is one of the best motivated ways to account for the observed neutrino masses. We discuss the expected sensitivity of future lepton collider experiments for probing such extensions. An interesting testable scenario is given by ``symmetry protected seesaw models'', which theoretically allow for sterile neutrino masses around the electroweak scale with up to order one mixings with the active (SM) neutrinos. In addition to indirect tests, e.g.\ via electroweak precision observables, sterile neutrinos with masses around the electroweak scale can also be probed by direct searches, e.g.\ via sterile neutrino decays at the Z pole, deviations from the SM cross section for four lepton final states at and beyond the WW threshold and via Higgs boson decays. We study the present bounds on sterile neutrino properties from LEP and LHC as well as the expected sensitivities of possible future lepton colliders such as ILC, CEPC and FCC-ee (TLEP).

\end{abstract}

\end{titlepage}

\setcounter{footnote}{0}

\section{Introduction}

The origin of the observed neutrino masses is one of the great open questions in particle physics. There are various ways to introduce massive neutrinos, which all require an extension of the particle content of the Standard Model (SM), or the introduction of effective operators which have to be generated at some higher energy scale by new physics involving additional particles. Currently, no experimental evidence exists to select between the various proposed extensions of the SM towards massive neutrinos.  

One of the best-motivated and most minimal extensions of the SM for providing neutrino masses consists in adding ``right-handed'' (often named ``sterile'') neutrinos to the SM degrees of freedom. Among the types of fermions within the SM, i.e.\ up-type quarks, down-type quarks, charged leptons and neutrinos, the neutrinos are the only type without a right-chiral counterpart. In Grand Unified Theories (GUTs) based on the gauge group SO(10), for instance, which has a left-right symmetric particle content, the ``right-handed'' neutrinos are therefore predicted. 

``Right-handed'' neutrinos would be SM-gauge singlets, and because of this they are often referred to as ``sterile''. Nevertheless, they can interact with the SM particles via their Yukawa couplings to the lepton doublets and the Higgs doublet. This coupling results in a Dirac-type mass for the neutrinos when the neutral component of the Higgs develops a non-zero vacuum expectation value. Furthermore, as gauge singlets, the sterile neutrinos can also have a mass term with their charge conjugates, i.e.\ a Majorana mass term. This leads to a mixing between the active and the sterile neutrinos. In the mass basis both, the light and the heavy eigenstates, couple to the $Z$ and the $W$ bosons.

With $n$ sterile neutrinos, the full neutrino mass matrix would be a $(3+n)\times(3+n)$ matrix. Out of the $3+n$ mass eigenstates, at least three have to be light, i.e.\ below about $0.5$ eV, in order to account for neutrinos oscillations, cosmological observations as well as constraints from neutrinoless double beta decay. The masses of the other (mainly sterile) neutrinos, as well as their couplings to the SM particles, are basically free parameters of the theory. 

Sterile neutrinos in various mass ranges have been discussed in the literature (see e.g.\ \cite{Abazajian:2012ys} for a review): For instance, sterile neutrinos with masses in the eV range could lead to effects in short distance neutrino oscillation experiments by introducing an additional mass squared difference. keV mass sterile neutrinos are candidates for ``warm'' dark matter, and very heavy sterile neutrinos around $M_\mathrm{GUT} \sim 10^{16}$ GeV are often predicted from GUTs. 

Here, we are interested in sterile neutrinos with masses around the electroweak (EW) scale, such that direct searches at present and future colliders are possible. Compared to our recent work \cite{Antusch:2014woa} where we assumed the sterile neutrinos to have masses sufficiently above the EW scale to test them via probes of non-unitarity of the effective leptonic mixing matrix, sterile neutrinos with EW masses can now be produced on-shell in particle collisions. In addition to indirect tests, e.g.\ via electroweak precision observables, they can now also be tested via, e.g., sterile neutrino decays at the Z pole, deviations from the SM cross section for four lepton final states at and beyond the W pair production  threshold, and via Higgs boson decays. 

In this paper we first study the present constraints on sterile neutrino properties in this mass range, including the whole relevant data from indirect tests (as in \cite{Antusch:2014woa}) as well as the present constraints from LEP and LHC on the above mentioned processes. Furthermore, we will provide first estimates for the expected sensitivities of future colliders such as the International Linear Collider (ILC), the Circular Electron Positron Collider (CEPC) and the electron-positron mode of the Future Circular Collider (FCC-ee, formerly known as TLEP), for testing EW sterile neutrinos and compare the prospects of direct and indirect searches. 

The paper is organized as follows: Section 2 contains a description of a minimal symmetry protected ``low scale'' type-I seesaw scenario, which we will use as benchmark framework for our analysis. In section 3 we derive the constraints from the present data, including ``direct'' and ``indirect'' searches for sterile neutrinos. In section 4 we present estimates for the sensitivity of future lepton colliders to sterile neutrino properties. Section 5 contains a discussion of our results and the conclusions.

\section{A symmetry protected ``low scale'' type-I seesaw scenario}

As described in the introduction, we investigate sterile neutrinos with masses around the EW scale. Such sterile neutrino masses can be realized in a ``natural'' way together with large (even {\cal O}(1)) Yukawa couplings to the lepton doublets and the Higgs doublet if there is a ``lepton-number-like'' symmetry which controls the size of the light neutrinos' masses, i.e.\ protects them from getting too large.\footnote{The term ``natural'' is understood here in the 't Hooft sense: Setting the masses of the light neutrinos to zero enhances the symmetry of the theory.} In fact neutrino masses in such scenarios are small when this protective symmetry is only slightly broken, in contrast to the usual seesaw mechanism where the smallness of the light neutrinos' masses comes from the heaviness of the sterile states. In our analysis we will focus on a minimal version of the symmetry protected scenario, where the experimentally observable effects stem from one pair of sterile neutrinos (having opposite charges under the protective symmetry).

\subsection{The scenario: Extension of the SM by EW scale sterile neutrinos}

To realize a low scale seesaw mechanism with a pair of sterile neutrinos $N_R^I$ $(I=1,2)$ without highly suppressed neutrino Yukawa couplings, we impose e.g.\ a ``lepton-number-like'' (global) $U(1)$ symmetry, where $N_R^1$ ($N_R^2)$ has the same (opposite) charge as the left-handed $SU(2)_L$ doublets $L^\alpha,\,\alpha=e,\mu,\tau$. Neutrino masses arise when this symmetry gets slightly broken, as e.g.\ in the so-called ``inverse'' \cite{Wyler:1982dd,Mohapatra:1986bd} or ``linear'' \cite{Malinsky:2005bi} variants of the type I seesaw mechanism (see also e.g.\ \cite{Shaposhnikov:2006nn,Kersten:2007vk,Gavela:2009cd}). The Lagrangian density in the symmetric limit is given by
\be
\mathscr{L} = \mathscr{L}_\mathrm{SM} -  \overline{N_R^1} M N^{2\,c}_R - y_{\nu_{\alpha}}\overline{N_{R}^1} \widetilde \phi^\dagger \, L^\alpha+\mathrm{H.c.}\;,
\label{eq:lagrange}
\ee
where $\mathscr{L}_\mathrm{SM}$ contains the usual SM field content and with $L^\alpha$ and $\phi$ being the lepton and Higgs doublets, respectively.\footnote{We remark that the ``lepton-number-like'' symmetry mentioned above is just an example and basically any symmetry leading to the above Lagrangian (with effects from possible additional terms being sufficiently suppressed) may be used.} 
The $y_{\nu_{\alpha}}$ are the complex-valued neutrino Yukawa couplings and the sterile neutrino mass parameter $M$ can be chosen real without loss of generality.

A third (or even more) sterile neutrinos may exist in addition, but we assume that it has (or they have) zero charge under the ``lepton-number-like'' symmetry such that in the symmetry limit they decouple from the other particles (since no Yukawa couplings to the lepton doublets are allowed and they also cannot mix with the other sterile states) and will be ignored. In this case, $M, y_{\nu_{e}},y_{\nu_{\mu}},y_{\nu_{\tau}}$ are the relevant parameters for studying the phenomenological consequences of a pair of EW scale sterile neutrinos.\footnote{ 
We note that in the specific case that indeed no further sterile neutrinos exist, i.e.\ only the two which form the pseudo-Dirac pair, then the lightest neutrino remains massless and there are correlations between the $y_{\nu_{e}},y_{\nu_{\mu}},y_{\nu_{\tau}}$ which depend on the elements of the light neutrino mixing matrix \cite{Gavela:2009cd}. Similarly, such correlations also arise in a special limit of the R-matrix parmeterisation \cite{Cely:2012bz}. In this study, we will not impose additional constraints of this type but consider the $y_{\nu_{e}},y_{\nu_{\mu}},y_{\nu_{\tau}}$ as independent parameters.
}
Only when the protective symmetry gets broken, all sterile neutrinos contribute to the generation of the light neutrino masses and  all three light neutrinos will obtain small masses. 
For phenomenological tests of the scenario at colliders or low energy precision experiments, the very small symmetry-breaking terms have negligible effects (see e.g.\ \cite{Kersten:2007vk,Gavela:2009cd}) and we will therefore study the limit where the symmetry is intact.

After electroweak symmetry breaking, one obtains the following $5 \times 5$ mass matrix of the (relevant) electrically neutral leptons:
\be
\mathscr{L}_{\rm mass} = -\frac{1}{2} \left(\begin{array}{c} \overline{\nu^c_{e_L}} \\ \overline{\nu^c_{\mu_L}} \\ \overline{\nu^c_{\tau_L}} \\ \overline{N_R^1} \\ \overline{N_R^2} \end{array}\right)^T\,
\left( \begin{array}{ccccc}  0 & 0 & 0 & m_e & 0 \\ 0 & 0 & 0 & m_\mu & 0 \\ 0 & 0 & 0 & m_\tau & 0 \\ m_e &  m_\mu & m_\tau & 0 & M \\ 0 & 0 & 0 & M & 0 \end{array}\right) \left(\begin{array}{c} \nu_{e_L}\\\nu_{\mu_L}\\\nu_{\tau_L}\\ \left(N_R^1\right)^c\\ \left(N_R^2\right)^c \end{array} \right) +\mathrm{H.c.}\,,
\label{eq:massmatrix}
\ee
with the Dirac masses $m_\alpha = y_{\nu_\alpha} v_\mathrm{EW}/\sqrt{2}$, where $y_{\nu_\alpha}$ are complex coupling constants and $v_\mathrm{EW}=246.22$ GeV. Note that in this limit of exact symmetry the right-handed neutrino $N_R^2$ does not couple to the SM leptons, and that the three lightest neutrinos are forced to be exactly massless. Diagonalising the mass matrix from eq.~(\ref{eq:massmatrix}), which we will denote by ${\cal M}$, with the unitary matrix $U$ yields the mass eigenstates: 
\be
U^T\, {\cal M}\, U = \text{Diag}\left(0,0,0,M,M\right)\,,
\ee
where we have neglected ${\cal O}(\theta^2)$ corrections to the masses of the heavy neutrinos. The (complex) active-sterile mixing parameters are defined as
\be
\theta_\alpha = \frac{y_{\nu_\alpha}^{*}}{\sqrt{2}}\frac{v_\mathrm{EW}}{M}\,,
\label{def:thetaa}
\ee
and the quantity
\be
\theta^2 = \sum_\alpha |\theta_\alpha|^2\,,
\label{def:theta}
\ee
such that to second order in the mixing parameter $\theta_\alpha$ the leptonic mixing matrix $U$ is unitary (cf. reference \cite{Antusch:2009gn}):
\be 
U = \left(\begin{array}{ccccc} 
{\cal N}_{e1}	& {\cal N}_{e2}	& {\cal N}_{e3}	& - \frac{\mathrm{i}}{\sqrt{2}}\, \theta_e & \frac{1}{\sqrt{2}} \theta_e 	\\ 
{\cal N}_{\mu 1}	& {\cal N}_{\mu 2}  	& {\cal N}_{\mu 3}  	& - \frac{\mathrm{i}}{\sqrt{2}}\theta_\mu & \frac{1}{\sqrt{2}} \theta_\mu  \\
{\cal N}_{\tau 1}	& {\cal N}_{\tau 2} 	& {\cal N}_{\tau 3} 	& - \frac{\mathrm{i}}{\sqrt{2}} \theta_\tau & \frac{1}{\sqrt{2}} \theta_\tau \\  
0	   	& 0		& 0	&  \frac{ \mathrm{i}}{\sqrt{2}} & \frac{1}{\sqrt{2}}\\
-\theta^{*}_e	   	& -\theta^{*}_\mu	& -\theta^{*}_\tau &\frac{-\mathrm{i}}{\sqrt{2}}(1-\tfrac{1}{2}\theta^2) & \frac{1}{\sqrt{2}}(1-\tfrac{1}{2}\theta^2)
\end{array}\right)\,.
\label{eq:mixingmatrix}
\ee
The elements of the non-unitary Pontecorvo--Maki--Nakagawa--Sakata (PMNS) matrix $\cal N$ can be written as \cite{Antusch:2009gn}
\be
{\cal N}_{\alpha i} = (\delta_{\alpha \beta} - \tfrac{1}{2} \theta_{\alpha}\theta_{\beta}^*)\,(U_\ell)_{\beta i}\,,
\label{eq:matrixN}
\ee
with $U_\ell$ being a unitary $3 \times 3$ matrix.

\subsection{Parameters}
As described above, the relevant new parameters of the scenario under consideration are the three complex Yukawa couplings $y_{\nu_e},y_{\nu_\mu},y_{\nu_\tau}$ and the mass $M$. Via eq.~(\ref{def:thetaa}), these parameters can be mapped onto the three (also complex) active-sterile mixing parameters $\theta_e,\theta_\mu,\theta_\tau$. 

Concerning physical processes, where the sterile neutrinos are very heavy compared to the experimental energy scale, they can be integrated out. The remaining effect is then given by the lepton-number conserving dimension six operator with coefficients
\be
c^{d=6}_{\alpha \beta} = \frac{y^*_{\nu_\alpha} y_{\nu_\beta}^{}}{M^2}\,,
\label{eq:O6pure}
\ee
which causes an effective non-unitarity of the leptonic mixing matrix via a contribution to the kinetic terms of the active neutrinos, 
as discussed e.g.\ in \cite{Antusch:2014woa}. 
The deviation of the PMNS matrix from unitarity, i.e.\ $\eps_{\alpha \beta} = ({\cal N N}^\dagger - \mathbbm{1})_{\alpha \beta}$, 
is obtained either from the coefficients in eq.~(\ref{eq:O6pure}) or from the definition of the mixing matrix ${\cal N}$ in eqs.~(\ref{eq:mixingmatrix}) and (\ref{eq:matrixN}) directly. To leading order in the mixing parameters the relation is given by
\be
\eps_{\alpha \beta} = - \frac{v^2_\mathrm{EW}}{2} c_{\alpha \beta}^{d=6} \equiv -\theta_\alpha^* \theta^{}_\beta \;,
\label{eq:epsYuk}
\ee
with the definition of the mixing $\theta_\alpha$ from eq.~(\ref{def:thetaa}). We summarise the parameters and the relevant mappings in tab.~\ref{tab:pars}.

\begin{table}
\begin{center}
\begin{tabular}{|l|c|c|c|}
\hline
 		& $y_{\nu_\alpha}$ 	& $\theta_\alpha$ & $\eps_{\alpha \beta}$ \\
\hline
$y_{\nu_\alpha} =$ 	& --		& ${\sqrt{2} M \over v_\mathrm{EW}}\theta^{*}_\alpha$ &  $-{\sqrt{2} M \over v_\mathrm{EW}} \, \eps_{\beta \alpha}/\sqrt{-\eps_{\beta \beta}}$ \\
\hline
$\theta_\alpha = $ & $\frac{v_\mathrm{EW}}{\sqrt{2} M} y^{*}_{\nu_\alpha}$ & -- &   $-\eps_{\beta \alpha}/\sqrt{-\eps_{\beta \beta}}$ \\
\hline
$\eps_{\alpha \beta} = $ & $ - \frac{v_\mathrm{EW}^2 y_{\nu_\alpha}^* y_{\nu_\beta}}{2 M^2}$  & $-\theta_\alpha^* \theta_\beta$ & --\\
\hline
\end{tabular}
\end{center}
\caption{
Relation between the sterile neutrino parameters $y_{\nu_\alpha}$ and $\theta_\alpha$, and the leptonic non-unitarity parameters $\eps_{\alpha \beta}$ used in ref.~\cite{Antusch:2014woa}.}
\label{tab:pars}
\end{table}

\subsection{Modification of the weak currents}
We can collect the left-handed neutrinos $\nu_\alpha$ and the charge conjugate right-handed fields $(N_R^1)^c,\,(N_R^2)^c$ into the column
\be
n = \left(\nu_{e_L},\nu_{\mu_L},\nu_{\tau_L},(N_R^1)^c,(N_R^2)^c\right)^T\,.
\ee
The mass eigenstates are given as
\be
\tilde n_j = \left(\nu_1,\nu_2,\nu_3,N_4,N_5\right)^T_j = U_{j \alpha}^{\dagger} n_\alpha\,.
\ee
Now we can write down the weak currents in the mass basis:
\bea
j_\mu^\pm & = & \sum\limits_{i=1}^5 \sum\limits_{\alpha=e,\mu,\tau}\frac{g}{\sqrt{2}} \bar \ell_\alpha\, \gamma_\mu\, P_L\, U_{\alpha i}\, \tilde n_i\, + \text{ H.c.}\,, \\
j_\mu^0 & = & \sum\limits_{i,j=1}^5 \sum\limits_{\alpha=e,\mu,\tau}\frac{g}{2\,c_W} \overline{\tilde n_j}\, U^\dagger_{j\alpha}\, \gamma_\mu\, P_L\, U_{\alpha i}\, \tilde n_i\,, 
\label{eq:weakcurrentmass}
\eea
where $g$ is the weak coupling constant, $c_W$ is the cosine of the Weinberg angle and $P_L = {1 \over 2}(1-\gamma^5)$ is the left-chiral projection operator. It is convenient to define the quantity
\be
\vartheta_{ij} = \sum_{\alpha=e,\mu,\tau} U^\dagger_{i\alpha}U_{\alpha j}^{}\,,
\ee
such that with $i\leq 3$ and $j=4,5$ we cobtain
\be
\vartheta_{i 4} = \sum\limits_{\alpha=e,\mu,\tau} (-\mathrm{i})\,{\cal N}_{i\alpha}^*\frac{\theta_\alpha}{\sqrt{2}}\,, \qquad \text{and} \qquad \vartheta_{i 5} = \sum\limits_{\alpha=e,\mu,\tau} {\cal  N}_{i\alpha}^*\frac{\theta_\alpha}{\sqrt{2}}\,,
\label{eq:heavylightmixing}
\ee
where the non-unitary PMNS matrix $\cal N$ was defined in eq.~(\ref{eq:matrixN}). 
The gauge couplings of the fermion current with two heavy neutrinos are proportional to $\vartheta_{jk}$ for $j,k=4,5$, which satisfy
\be
|\vartheta_{jk}| =  {1 \over 2}\, \theta^2\,,
\label{def:eta55}
\ee
with $\theta^2$ defined in eq.~(\ref{def:theta}).
With the above definitions, the weak currents involving the heavy neutrinos in the mass basis can be written compactly as
\bea
j_\mu^\pm & \supset &  \frac{g}{2} \, \theta_\alpha \, \bar \ell_\alpha \, \gamma_\mu P_L \left(-\mathrm{i} N_4 + N_5 \right) + \text{H.c.} \,, \label{eq:weakcurrent1}\\
j_\mu^0 & = & \frac{g}{2\,c_W} \sum\limits_{i,j=1}^5 \vartheta_{ij} \overline{ \tilde n_i} \gamma_\mu P_L \tilde n_j\,.
\label{eq:weakcurrent2}
\eea
Analogously we can express the Yukawa part of the Lagrangian density in the mass basis, 
\begin{align}
\sum\limits_\alpha y_{\nu_\alpha}\overline{N^1_R} \widetilde \phi^\dagger L^\alpha  + \text{H.c.}\supset  \sum\limits_\alpha y_{\nu_\alpha}\sum\limits_{i,j} \overline{\tilde{n}^c}_j U_{j 4}^{T}\,\phi^0\, U_{\alpha i}^{} \tilde n_i^{} +\text{ H.c.} \notag \\
 = \frac{\sqrt{2}\,M}{v_\mathrm{EW}} \left[\sum\limits_{i=1}^3 \left(\vartheta_{i4}^* \overline{N_4^c}+ \vartheta_{i5}^*\overline{N^c_5}\right) \phi^0 \nu_i + \sum\limits_{j=4,5} \vartheta_{jj}^* \overline{N^c_j} \phi^0 N_j \right] +\text{ H.c.}  \,,
\label{eq:Lykawa}
\end{align}
with being $\phi^0$ the neutral component of the Higgs doublet.

\subsection{Decay rates involving heavy sterile neutrinos}
With the weak currents in eq.~(\ref{eq:weakcurrent1}) and (\ref{eq:weakcurrent2}) and the Yukawa terms in eq.~(\ref{eq:Lykawa}), the heavy neutrinos $N_4$ and $N_5$ couple to the weak gauge bosons and the Higgs boson, respectively. They can either be produced in decays from gauge and Higgs bosons, or decay into leptons and bosons, depending on which process is kinematically allowed. 

First, we consider the case of $M < m_W,m_Z, m_h$, which yields the following decay channels: 
\be
W^+ \to \bar \ell \, N\,, \qquad W^- \to \ell \, \bar N\,, \qquad Z \to \bar N\, \nu\,,\qquad Z \to \bar N\, N\,,\qquad h \to \bar N\,\nu \,.
\ee
We have suppressed here the indices of the neutrino mass eigenstates and of the leptons. $W,\,Z$ are the weak gauge bosons, $h$ is the SM Higgs boson and $\ell = e,\,\mu,\,\tau$ denote the charged leptons. Note, that for the $Z$ and Higgs boson decays also the Hermitean conjugate processes have to be taken into account. 
Neglecting the masses of the light neutrinos and charged leptons the corresponding decay rates for $i=1,2,3,\, j,k=4,5$ are given by 
\bea
\Gamma(W^- \to \overline{N}_j \ell_\alpha^-) & = & \frac{|\theta_\alpha|^2}{2} \frac{G_F\,m_W^3}{6\sqrt{2}\pi}\Pi_{(1+1)}(\mu_W)\,, \label{eq:W->lN} \\
\Gamma(W^+ \to N_j \overline{\ell_\alpha^-}) & = & \Gamma(W^- \to \overline{N}_j \ell_\alpha^-) \\
\Gamma(Z \to \bar{\nu}_i N_j) & = & |\vartheta_{ij}|^2 \frac{G_F\,m_Z^3}{6\sqrt{2}\pi}\Pi_{(1+1)}(\mu_Z)\,, \label{eq:Z->nuN}\\
\Gamma(Z \to \overline N_j N_k) & = & |\vartheta_{jk}|^2 \frac{G_F\,m_Z^3}{6\sqrt{2}\pi}\Pi_{(2)}(\mu_Z)\,, \label{eq:Z->NN}\\
\Gamma(h \to \bar \nu_{i} N_j) & = & \frac{m_h\, \left|\vartheta_{ij}\right|^2 M^2}{16\, \pi\,v_\mathrm{EW}^2}\left(1-\mu_h^2\right)^2\,, \label{eq:htoNnu}
\eea
where we introduced $\mu_X = M/m_X$, $G_F$ is the Fermi constant, and the kinematic factors are
\bea
\Pi_{(1+1)}(\mu_X) & = & {1\over 2}\left(1-\mu_X^2\right)^2 \left(2+\mu_X^2\right)\,, \label{eq:kinfac1}\\
\Pi_{(2)}(\mu_X) & = & {1\over 2}\left(1-{\mu_X^2 \over 4}\right)^2 \left(2+\mu_X^2\right)\,.
\eea
We note that the decay rates of $Z$ to $\overline{N}$ are the same as the ones to $N$, e.g.\ $\Gamma(Z \to \nu_{i} \overline{N}_j) = \Gamma(Z \to \bar{\nu}_{i} N_j)$. Both processes have to be taken into account when calculating $R_{inv}$, as will be discussed below. 

To obtain the total Higgs decay rate into neutrinos, we observe that to leading order in the active-sterile mixing parameters
\be
\sum\limits_{i=1}^3\left(|\vartheta_{i4}|^2+|\vartheta_{i5}|^2\right) = |\theta_e|^2 + |\theta_\mu|^2 + |\theta_\tau|^2\,. 
\label{eq:etatheta}
\ee
Therefore, by summing eq.~(\ref{eq:htoNnu}) over $j=4,5$ and $i=1,2,3$, and including the Hermitian conjugate process, we obtain to leading order in the mixing parameters
\be
\Gamma(h \to \nu N) = \frac{m_h\, \theta^2\,M^2}{8\, \pi\,v_\mathrm{EW}^2}\left(1-\mu_h^2\right)^2\,.
\label{eq:higgstoneutrinos}
\ee
The complementary processes which are kinematically available for $M>m_W,m_Z,m_h$, namely the corresponding decay rates for the heavy neutrinos, i.e. $j=4,5$, are
\be
N_j \to W\, \ell_\alpha\,, \qquad N_j \to Z \nu_{i}\,, \qquad N_j \to h \nu_{i}\,.
\ee
The corresponding decay rates for $i=1,2,3,\,j=4,5$ are given by
\bea
\Gamma(N_j \to W\, \ell_\alpha) & = & \frac{|\theta_\alpha|^2}{2} \frac{G_F\,M^3}{4\sqrt{2}\pi}\Pi_{(1+1)}(\mu_W^{-1})\,, \\
\Gamma(N_j \to Z\,\nu_i) & = & |\vartheta_{ij}|^2 \frac{G_F\,M^3}{4\sqrt{2}\pi}\Pi_{(1+1)}(\mu_Z^{-1})\,, \\
\Gamma(N_j \to h\,\nu_i) & = & |\vartheta_{ij}|^2 \frac{M^3}{16\,\pi\,v_\mathrm{EW}^2}\left(1-\mu_h^{-2}\right)^2\,.
\eea
We use the following parametric values \cite{Beringer:1900zz}:
\begin{center}
\begin{tabular}{|l|c|c|c|c|}
\hline
Parameter & $m_Z$ [GeV] & $m_W$ [GeV] & $m_h$ [GeV] & $G_F$ [GeV$^{-2}$] \\
\hline
\hline
Value & 91.1875 & 80.358 & 126.0 & 1.1663787$\times 10^{-5}$ \\
\hline
\end{tabular}
\end{center}

\section{Present Constraints}

Before we study the sensitivities of future colliders, we discuss the constraints on sterile neutrino properties from the currently available experimental data. We start with ``indirect'' constraints from precision tests of the SM and then turn to ``direct'' tests focusing on sterile neutrino decays at the Z pole, deviations from the SM cross section for four lepton final states at and beyond the WW threshold and Higgs boson decays.

\subsection{``Indirect'' constraints from precision tests of the SM}
\label{sec:indirectconstraints}
We consider the mass of the heavy neutrinos, $M$ (note that we have only one mass scale here due to the protective symmetry), to be in the range from  $\sim$ 10 GeV to $\sim 250$ GeV. 
In the presence of the sterile neutrinos the theory predictions for various precision observables get modified. In this subsection we extend the analysis of our recent work \cite{Antusch:2014woa}, where we assumed the heavy neutrinos to have masses sufficiently above the EW scale to test them via probes of non-unitarity of the effective leptonic mixing matrix, to masses around the EW scale. 
A discussion of constraints on sterile neutrinos in the mass range below 10 GeV can be found in ref.~\cite{Drewes:2015iva}.

For various observables, where the experiments are performed at energies much below $M$, the results from \cite{Antusch:2014woa} still apply (since the effective theory treatment is still applicable) and we can simply translate them into constraints on the sterile mixing parameters using table \ref{tab:pars}. We will mainly revisit the observables where there are changes due to $M$ around the EW scale, such as the electroweak precision observables (EWPOs) at colliders as well as the rare charged lepton flavour violating (LFV) decays.

\subsubsection*{Effects on the Fermi constant}
The Fermi constant $G_F$ is measured from muon decays which get modified due to the effects of the sterile neutrinos in the charged current interactions. Denoting the Fermi constant extracted from muon decays as $G_\mu$, we obtain the relation
(at tree-level and to leading order in the mixing parameters) 
\be
 G_\mu^2 = G_F^2(1-|\theta_e|^2)(1-|\theta_{\mu}|^2)\;.
\label{eq:GF}
\ee
This has consequences for the theory predictions of many precision observables. 

\subsubsection*{Electroweak precision observables}
The tree level relation between $s_W = \sin\theta_W$, $G_F$ and $\alpha$ is given by
\be
\sw c^2_W  = \frac{\alpha(m_Z) \pi}{\sqrt{2}  G_F m_Z^2}\,,
\label{eq:treelevel}
\ee
which yields the following theory prediction for $s_{W}^2$, which is modified with respect to the SM due to eq.~(\ref{eq:GF}):
\be
s_{W}^2 = \frac{1}{2}
\left[1-\sqrt{1-\frac{2\sqrt{2}\alpha \pi}{G_\mu m_Z^2} \sqrt{(1-|\theta_e|^2)(1-|\theta_{\mu}|^2)}}
\right]\,.
\label{eq:seff}
\ee
Furthermore, together with the tree-level relation $m_Z^2 c_W^2 = m_W^2$, the theory prediction for the $W$ boson mass is modified to
\be
m_W^2 = [m_W^2]_{\rm SM} \left[\sqrt{(1-|\theta_e|^2)(1-|\theta_{\mu}|^2)}
\frac{[s_W^2]_{\rm SM}}{s_W^2}
\right]\,,    \label{eq:mW}
\ee
with the weak mixing angle $s_W$ from eq.~(\ref{eq:seff}).

\subsubsection*{$Z$ boson decay parameters}
The modification of the Fermi constant in eq.~(\ref{eq:GF}) also modifies the tree level decay rate of the $Z$ boson into fermions $f \bar f$. For $f \neq \nu$ we have 
\be
\Gamma_{Z\to ff} =  N_c^f\frac{ G_\mu M_Z^3}{6 \sqrt{2} \pi} \frac{\left(g_{A,f}^2 + g_{V,f}^2\right)}{\sqrt{(1-|\theta_\mu|^2)(1-|\theta_e|^2)}}\,,
\label{eq:gzvis}
\ee
with $N_c$ being the colour factor and $g_{V,f},\,g_{A,f}$ the vector and axial vector coupling
\be
g_{V,f} = T_3^f - 2 Q_f s_W^2\,, \qquad g_{A,f} = T_3^f\,,
\label{eq:leptoncoupling}
\ee
with the third component of the isospin $T_3^f$ and the electric charge $Q_f$. Through $G_\mu$ and $s_W$, the decay rate $\Gamma_{ff}$ is affected by the modification of the light neutrino couplings.
An observable which is very sensitive to the modifications due to the mixing parameters is the decay rate of the $Z$ boson into two light neutrinos, i.e.\ with $i,j\leq3$ and to leading order in the active-sterile mixing parameters:
\be
\sum\limits_{i,j=1}^3\Gamma_{Z\to \nu_i\nu_j} = \sum\limits_{\alpha,\beta} (\delta_{\alpha \beta} - \theta_{\alpha}^* \theta_\beta^{})^2 \times \Gamma_{Z\to \nu,\,\rm SM}\times \left[(1-\theta_e|^2)(1-|\theta_{\mu}|^2)\right]^{-{1\over 2}}\,,  
\label{eq:Znulight}
\ee
where $\Gamma_{Z\to \nu,\,\rm SM}=G_\mu m_Z^3/(6 \sqrt{2} \pi)$ is the decay width for $Z\to \bar \nu_\alpha \nu_\alpha$ for a specific flavour, in the SM. 

We now turn to the hadronic pole cross section $\sigma_{had}^0$ and the invisible decay rate $R_{inv}$ of the $Z$ boson, defined as
\be
\sigma_{had}^0 =  \frac{12 \pi}{M_Z^2} \frac{\Gamma_{Z\to ee}\Gamma_{Z\to had}}{\Gamma_Z^2}\,,  \qquad
R_{inv} = \frac{\Gamma_{Z\to  inv}}{\Gamma_{Z\to \ell\ell}}\,,
\ee
where $\Gamma_{Z\to inv}$ is the invisible partial decay width and $\Gamma_{Z\to had}$ is the sum over all the hadronic partial decay widths of the $Z$ boson and $\Gamma_Z$ is the total $Z$ decay width.  To leading order in $\theta_\alpha$ we obtain the following expression

\bea
\sigma_{had}^0 & = & \left[\sigma_{had}^0 \right]_{\rm SM}(1 + 0.27 \theta^2\left(1 + c_\sigma \Pi_{(1+1)}(\mu_Z) \right) - 0.02\left(|\theta_e|^2+|\theta_\mu|^2\right)  \,, \label{eq:sigmah} \\
R_{inv} & = & \left[R_{inv} \right]_{\rm SM}\left(1 - {2\over 3}\theta^2\left(1 + c_R \, \Pi_{(1+1)}  (\mu_Z) \right)\right) - 0.09\left(|\theta_e|^2+|\theta_\mu|^2\right)\,, \label{eq:Rinv}
\eea
where we used the values for the parameters $c_\sigma=-0.82$ and $c_R=-0.67$. If the heavy neutrinos would not decay inside the detector, the $c_{\sigma,R}$ would be $-1$.
For estimating the parameters $c_{\sigma,R}$ we have assumed that the heavy neutrinos decay within the detector (as will be the case for most of the considered parameter space) and that the kinematically available SM fermions are massless. Furthermore, following \cite{Decamp:1990ky}, we assumed that all processes where the heavy neutrinos decay into hadrons (and a light neutrino) are counted as hadronic events, whereas all leptonic and semileptonic $N$ decays are rejected by the event selection filters. This estimate is sufficient for the discussion in this paper, but should be replaced by a more accurate treatment at latest when a signal is found.

It is often stated that for heavy neutrino masses much smaller than $m_Z$ unitarity of the PMNS matrix is effectively recovered and that therefore the prediction for the invisible $Z$ decay rate coincides again with the SM one (with $N_\nu=3$). However, as long as the heavy neutrino is heavier than the muon, there is in any case a dependency on the mixing parameter combination $|\theta_{e}|^2+|\theta_{\mu}|^2$ due to the use of the Fermi constant as input parameter.

Analogously to the above discussions for $R_{inv}$ and $\sigma_{had}^0$, we also include the pseudo-observables $R_\ell,R_b$ and $R_c$: 
\bea
R_\ell & = & \left[R_\ell \right]_{\rm SM}(1+0.15 (|\theta_e|^2+|\theta_\mu|^2) -0.07\theta^2\Pi_{(1+1)}(\mu_Z))\,,   \\
R_b & = & \left[R_b \right]_{\rm SM}(1 - 0.03 (|\theta_e|^2+|\theta_\mu|^2) - 0.001\theta^2\Pi_{(1+1)}(\mu_Z))\,, \\
R_c & = & \left[R_c \right]_{\rm SM}(1 + 0.06 (|\theta_e|^2+|\theta_\mu|^2) - 0.0003\theta^2\Pi_{(1+1)}(\mu_Z))\,. 
\eea
We assume that the sterile neutrino decays do not significantly affect the experimental determination of $m_W$ and the effective weak mixing angle (such that they are only sensitive to $|\theta_e|^2+|\theta_\mu|^2$ due to $G_F$ as given in eqs.~(\ref{eq:seff}) and (\ref{eq:mW})).
The SM predictions and experimental values for the EWPOs are taken from ref.~\cite{Baak:2014ora}.

\subsubsection*{Lepton universality observables}
The lepton universality observables considered here are defined as ratios of decay rates: $R^X_{\alpha\beta}=\Gamma^X_\alpha/\Gamma^X_\beta$, where $\Gamma^X_\alpha$ denotes a  decay width including a charged lepton $\ell_\alpha$ and a neutrino. They are defined such that in the SM $R^X_{\alpha\beta}=1$ holds for all $\alpha,\beta,X$, as a consequence of lepton universality. 
We include here constraints from $\pi$, $\mu$, $\tau$ and $K$ decays which stem from experiments at comparatively low energy and which are currently dominating the constraints. This allows us to use the results from \cite{Antusch:2014woa} for these processes, translating the parameters using table \ref{tab:pars}. 
Constraints on sterile neutrinos from lepton universality tests have also been studied recently in refs.~\cite{Abada:2012mc,Abada:2013aba}.
We note that in contrast to \cite{Antusch:2014woa} we are not including $W$ decays here, which however only had a negligible impact on results of the fit in \cite{Antusch:2014woa}. 
The active-sterile mixing between the left-handed and right-handed neutrinos leads to modified theory predictions of the form 
\be
R^X_{\alpha \beta} = 1 - {1 \over 2}\left(|\theta_{\alpha}|^2-|\theta_{\beta}|^2\right)\,,
\ee
which thus allows to probe differences between the $\theta_{\alpha}$. We display the present experimental constraints on the universality observables in tab.~\ref{tab:universality}.

\begin{table}
\begin{center}
$\begin{array}{|c|c|c|c|c|c|c|} 
\hline
 & R_{\mu e}^\ell & R_{\tau \mu}^\ell & R_{\mu e}^\pi & R_{\tau \mu}^\pi & R^K_{\tau \mu} & R^K_{\tau e} \\
\hline\hline
\mathrm{Process} &  \frac{\Gamma (\tau \to \nu_\tau \mu \bar{\nu}_\mu )}{\Gamma (\tau \to \nu_\tau e \bar{\nu}_e )} & \frac{\Gamma (\tau \to \nu_\tau e \bar{\nu}_e )}{\Gamma (\mu \to \nu_\mu e \bar{\nu}_e )} & \frac{\Gamma (\pi \to \mu \bar{\nu}_\mu )}{\Gamma (\pi \to e \bar{\nu}_e )} & \frac{\Gamma (\tau \to \nu_\tau \pi)}{\Gamma (\pi \to \mu \bar{\nu}_\mu )} & \frac{\Gamma(\tau \to K \nu_\tau )}{\Gamma(K \to \mu \bar{\nu}_\mu)} & \frac{\Gamma(\tau \to K \nu_\tau )}{\Gamma(K \to e \bar{\nu}_e)} \\
\hline
\mathrm{Bound} & 1.0018(14) & 1.0006(21) & 1.0021(16) & 0.9956(31) & 0.9852(72) & 1.018(42) \\
\hline
\end{array}$
\end{center}
\caption{Tests of lepton universality used in our global fit. Experimental results are taken from ref.~\cite{Amhis:2012bh}.}
\label{tab:universality}
\end{table}

\subsubsection*{Rare flavour-violating charged lepton decays}

The decay rate for lepton flavour violating charge lepton decays $\ell_\rho\to \ell_\sigma \gamma$ are given by
\be
\Gamma_{\ell_\rho\to \ell_\sigma \gamma} = \frac{ \alpha G_\mu^2 m_\rho^5}{2048 \pi^4}|\sum\limits_{k=1}^5 U_{\rho k}^{} U^\dagger_{k \sigma} F(x_k) |^2 \,,
\label{eq:raredecay}
\ee
where terms $\sim {\cal O}((m_{\ell_\sigma}/m_{\ell_\rho})^2)$ are neglected and where $F(x_k)$ is a loop-function which depends on the mass ratio $x_k = |m_{\nu_k}/M_W|$:
\be
F(x) =  \frac{10-43x+78x^2-49x^3+4x^4 +18x^3 \ln x}{3(1-x)^4}\,, \qquad F(0) = {10 \over 3}\,.
\label{eq:loopfunction}
\ee
The $m_{\nu_k}$ are the mass eigenvalues of the light and heavy neutrinos, and we shall approximate them with $0$ and $M$, respectively. By using the unitarity of the neutrino mixing matrix $U$ (up to second order in $\theta_\alpha$), we can write
\bea
\sum\limits_{k=1}^5 U_{\rho k}^{} U^\dagger_{k \sigma} F(x_k) & = & \sum\limits_{k=1}^3 U_{\rho k}^{} U^\dagger_{k \sigma} F(0) + \sum\limits_{k=4}^5 U_{\rho k}^{} U^\dagger_{k \sigma} F(x_M) \notag \\
& = & -\theta_\rho^{}\theta_\sigma^*\left[F(0) - F(x_M)\right]\,,
\eea
with $x_M = M/m_W$. Plugging this into eq.~(\ref{eq:raredecay}), the branching ratio for the process $\ell_\rho \to \ell_\sigma \gamma$ can be expressed as
\be
Br_{\rho\sigma} = Br(\ell_\rho \to \nu_\rho \bar \nu_\sigma \ell_\sigma) \frac{100 \alpha}{96 \pi}|\theta_{\rho}|^2|\theta_\sigma|^2\, \left[1-{3 \over 10}F\left(x_M\right)\right]^2\,.
\ee
Notice, that with $M \to 0$ unitarity is restored, as it should be. Furthermore, we remark that in the limit $M \gg m_Z$, the above expression seems to differ from the corresponding one in ref.~\cite{Antusch:2014woa}, where the low energy effective theory was considered. Note however that ``decoupling'' automatically implies $\theta \to 0$, such that in both frameworks the low energy effects of the extra sterile neutrinos disappear as they should. The present experimental constraints on rare charged lepton decays are summarised in tab.~\ref{tab:rcld}.

\begin{table}
\begin{center}
\begin{tabular}{|c|c|c|}
\hline
Process & Prediction with active-sterile neutrino mixing & 90 \% C.L. bound \\
\hline\hline
$Br(\mu  \to e \gamma)$ & $2.4\times10^{-3}\left(1- 0.3\, F(x_M)\right)|\theta_\mu^*\theta_e^{}|^2$ & 5.7 $\times 10^{-13}$ \\
$Br(\tau \to e \gamma)$ & $4.3\times10^{-4}\left(1- 0.3\, F(x_M)\right)|\theta_\tau^*\theta_e^{}|^2$ & 1.5 $\times 10^{-8}$ \\
$Br(\tau \to \mu \gamma)$ & $4.1\times10^{-4}\left(1- 0.3\, F(x_M)\right)|\theta_\tau^*\theta_\mu^{}|^2$ & 1.8 $\times 10^{-8}$ \\
\hline
\end{tabular}
\end{center}
\caption{Present bounds on the charged lepton flavour violating processes $\ell_\alpha \to \ell_\beta \gamma$ and predictions in the presence of sterile neutrinos. The experimental bounds on $\mu \to e \gamma$ are from the MEG collaboration~\cite{Adam:2013mnn}, the ones on $\tau$ decays are taken from ref.~\cite{Blankenburg:2012ex}. The function $F(x)$ is defined in eq.~(\ref{eq:loopfunction}).}
\label{tab:rcld}
\end{table}

\subsubsection*{Other precision constraints} 

In addition to the ``indirect'' tests mentioned above, we also include the constraints from the NuTeV experiment, from CKM unitarity tests and from low energy measurements of $s_W^2$. To calculate these additional constraints we follow the analysis of \cite{Antusch:2014woa}, translating the parameters using table \ref{tab:pars}. The modified theory predictions for the EWPOs, NuTeV observables and low energy measurements of the weak mixing angle due to active-sterile neutrino mixing are summarized in tab.~\ref{tab:indirect2}. The constraints from CKM unitarity tests are discussed in ref.~\cite{Antusch:2014woa} in sec.\ 3.2.3.

\begin{table}[h!]
\begin{center}
\begin{tabular}{|l|c|c|c|}
\hline
Prediction with heavy neutrinos & Prediction in the SM & Experiment \\
\hline\hline
$[m_W]_{\rm SM}(1+0.11 (|\theta_e|^2+|\theta_\mu|^2))$ 		& 80.358(8) GeV & 80.385(15) GeV  \\
$[\Gamma_{\rm lept}]_{\rm SM}(1+0.59(|\theta_e|^2+|\theta_\mu|^2))$  	& 83.966(12) MeV & 83.984(86) MeV \\
$[(s_{W,\mathrm{eff}}^{\ell,\mathrm{lep}})^2]_{\rm SM}(1-0.71(|\theta_e|^2+|\theta_\mu|^2))$	& 0.23150(1) & 0.23113(21)  \\
$[(s_{W,\mathrm{eff}}^{\ell,\mathrm{had}})^2]_{\rm SM}(1-0.71(|\theta_e|^2+|\theta_\mu|^2))$  & 0.23150(1)	& 0.23222(27) \\
\hline
$\left[R_\nu\right]_{\rm SM}(1 + 0.3|\theta_{e}|^2 - 1.7 |\theta_{\mu}|^2)$  & 0.3950(3) & 0.3933(15) \\
$\left[R_{\bar\nu}\right]_{\rm SM}(1 + 0.1|\theta_{e}|^2 - 1.9 |\theta_{\mu}|^2)$ & 0.4066(4) & 0.4034(28) \\
\hline
$\left[ Q^{55,78}_W\right]_{\rm SM}(1-0.48 (|\theta_e|^2 + |\theta_\mu|^2))$ & -73.20(35) & -72.06(44) \\
$\left[ Q^p_W\right]_{\rm SM}(1+9.1(|\theta_e|^2 + |\theta_\mu|^2))$ & 0.0710(7) & 0.064(12) \\
$\left[ A_{LR}^{ee}\right]_{\rm SM}(1+15.1(|\theta_e|^2 + |\theta_\mu|^2))$ & 1.520(24)$\times10^{-7}$ & 1.31(17)$\times 10^{-7}$ \\
\hline
\end{tabular}
\end{center}
\caption{
Experimental results, SM predictions and the modification in the presence of sterile neutrinos for $m_W$, the effective weak mixing angle, the NuTeV observables and for the low energy measurements of the weak mixing angle. The SM predictions and experimental values for the EWPOs are taken from ref.~\cite{Baak:2014ora}. 
The values of $(s_{W,\mathrm{eff}}^{\ell,\mathrm{lep}})^2$ and $(s_{W,\mathrm{eff}}^{\ell,\mathrm{had}})^2$ are taken from Ref.~\cite{Ferroglia:2012ir}.
The NuTeV results on deep inelastic scattering of neutrinos and anti-neutrinos on nuclear matter has been taken from ref.~\cite{Bentz:2009yy}. The theory uncertainty stems from $s_W^2$. The results on $Q^p_W$ are from Ref.~\cite{Nuruzzaman:2013bwa}. For $[A_{LR}^{ee}]_{SM}$ we used $s_W^2(M_Z)=0.2315$, and its error is dominated by the uncertainty of the radiative QED correction factors.
}
\label{tab:indirect2}
\end{table}

\subsubsection{Present constraints from ``indirect'' tests: Global fit results}
To obtain the present constraints on sterile neutrino extensions of the SM from precision observables, we perform a Markov Chain Monte Carlo (MCMC) fit for the three Yukawa couplings and mixing parameters, $y_{\nu_\alpha}$ and $\theta_\alpha$, for 10 GeV $\leq M \leq$ 250 GeV, and extract the highest posterior probability density (HPD) intervals at 90\% confidence level (CL). We use the experimental constraints discussed in the previous section, which are essentially based on the observables also used in ref.~\cite{Antusch:2014woa} unless stated otherwise in the text, adapted to sterile neutrino scenarios with M in the EW range.

\begin{figure}
\begin{minipage}{0.49\textwidth}
\begin{center}
\includegraphics[scale=0.45]{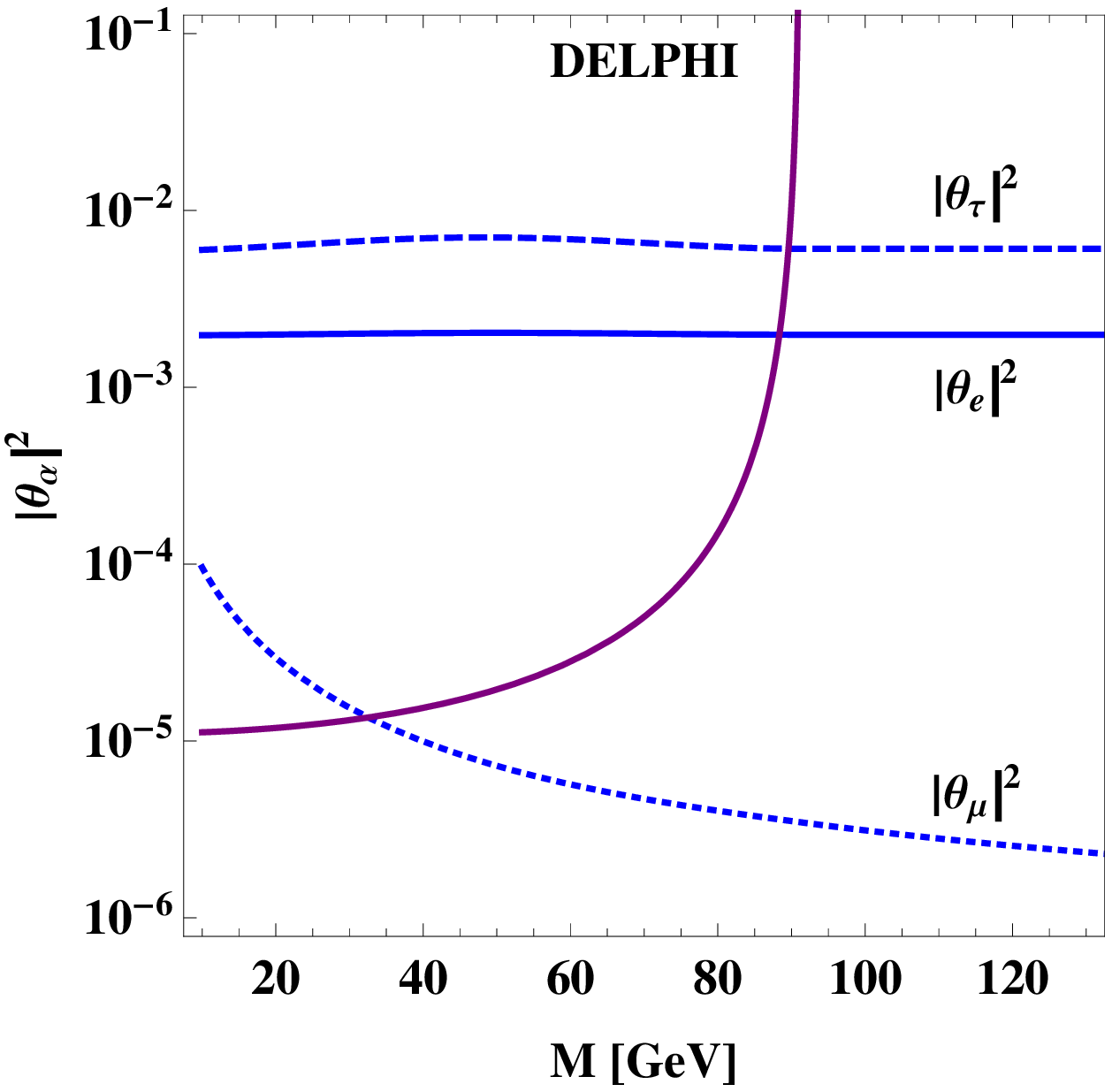}
\end{center}
\end{minipage}
\begin{minipage}{0.49\textwidth}
\begin{center}
\includegraphics[scale=0.45]{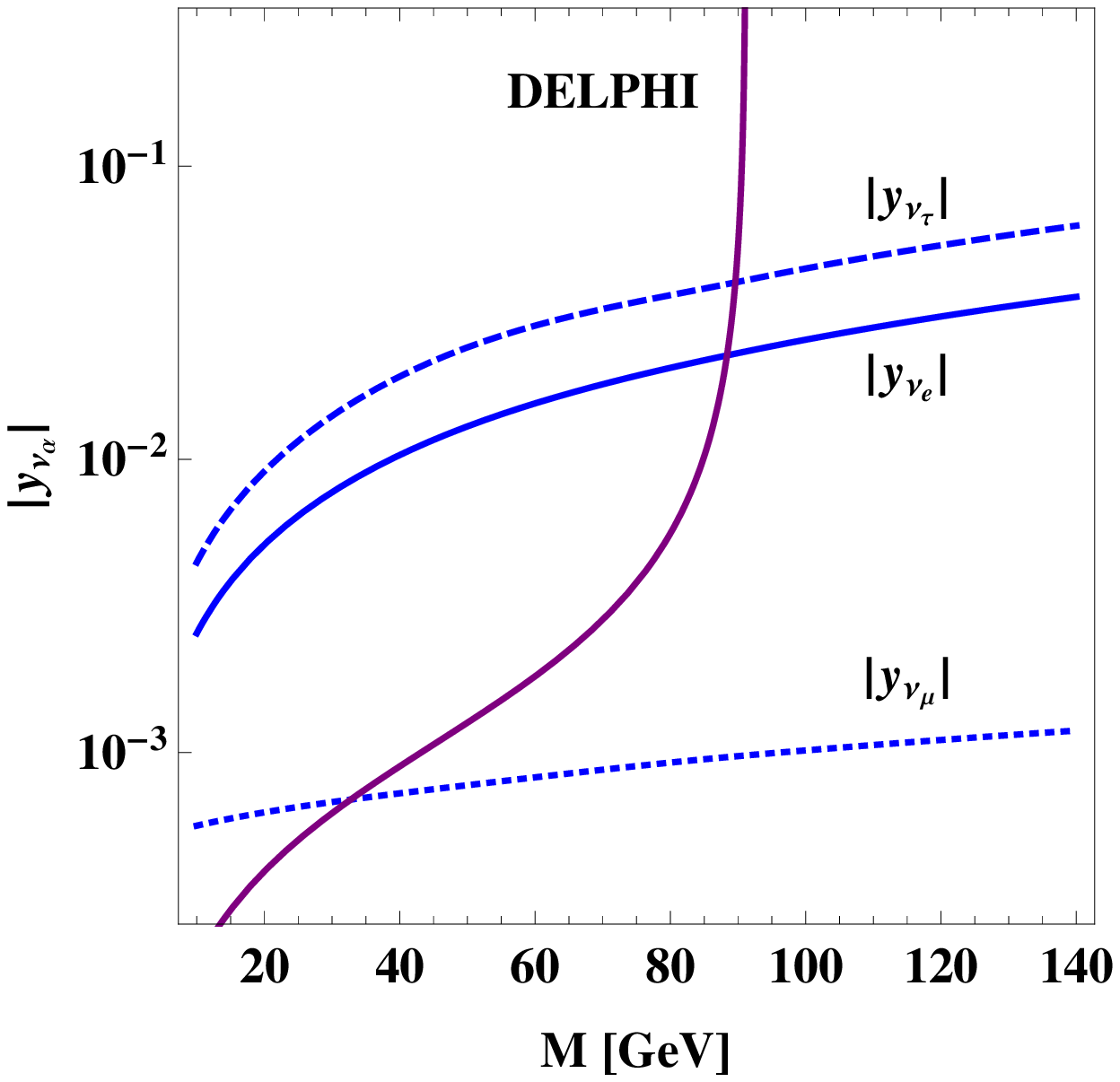}
\end{center}
\end{minipage}
\caption{Upper limits on the active-sterile neutrino mixing parameters from ``indirect'' tests at 90\% CL. The {\it left} panel shows the mixing parameters $\theta_\alpha,\, \alpha=e,\mu,\tau$, the {\it right} panel shows the Yukawa couplings $y_{\nu_\alpha}$. The purple line represents the direct search constraints on the parameter space from Delphi \cite{Abreu:1996pa}.}
\label{fig:yukawa_constraints}
\end{figure}

We display the resulting upper bounds for the parameters $|\theta_\alpha |$ and $|y_{\nu_\alpha}|$,  with $\alpha=e,\mu,\tau$, in fig.~\ref{fig:yukawa_constraints}. For comparison, we also show the exclusion limit  from the direct searches of the LEP-I experiment Delphi, which will be discussed below. 

We find, in agreement with the results in refs.~\cite{Antusch:2014woa,Basso:2013jka}, that for $|y_{\nu_e}|$ there is also a non-zero lower bound at 90\% confidence level. In the following, however, we will only use the upper bound on $|y_{\nu_e}|$ as constraint. The best fit value for $|y_{\nu_\mu}|$ is zero and the uncertainty on $|y_{\nu_\tau}|$ is much larger than that of the other two parameters.

\subsection{Present constraints from ``direct'' searches}

We now turn to the current constraints from ``direct'' searches, i.e.\ via sterile neutrino decays at the Z pole, deviations from the SM cross section for four lepton final states at and beyond the WW threshold, and via Higgs boson decays. As we will discuss below, especially the first two tests provide the strongest constraints for specific mass ranges (below $M \sim 150$  GeV), whereas the ``indirect'' tests are more sensitive for larger $M$. In the next section we will estimate the sensitivity improvements which could be possible at envisioned future colliders. 

\subsubsection{Search for sterile neutrinos produced in $Z$ boson decays}
\label{sec:delphi}
\begin{figure}
\begin{center}
\includegraphics[scale=0.67]{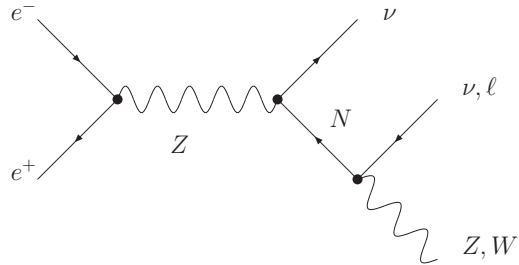}
\end{center}
\caption{Feynman diagram dominating the production of sterile neutrinos at the $Z$ pole.}
\label{fig:nzpole}
\end{figure}

The LEP-I collaborations Delphi \cite{Abreu:1996pa}, Opal \cite{Akrawy:1990zq}, Aleph \cite{Decamp:1991uy} and L3 \cite{Adriani:1993gk} have performed analyses searching for ``heavy neutral leptons'' -- or sterile neutrinos --  at the $Z$-pole. 
The Feynman-diagram for sterile neutrino production which is dominant at the $Z$ pole is shown in fig.~\ref{fig:nzpole}.
The results of the LEP collaborations can be expressed as an upper limit on the branching ratio for $Z$ bosons decaying into a light and a heavy neutrino. It can be used to constrain the sterile neutrino parameters as we now discuss: 

The strongest bound on the branching ratio for the processes $Z \to \nu \, N$ comes from the Delphi collaboration. It is given at 95\% C.L.\ as
\be
Br(Z \to \nu \, N) < 1.3 \times 10^{-6}\,,
\label{eq:eetonuN}
\ee
which includes the processes $Z \to \bar \nu_{i\leq 3} \,N_j,\, j=4,5$ and the Hermitean conjugate processes. 

With the expression for the corresponding decay rate in eq.~(\ref{eq:Z->nuN}), the experimental upper bound from eq.~(\ref{eq:eetonuN}) can be used to put upper bounds on the sum over all the active-sterile mixing parameters:
\be
\theta^2 \leq \frac{1.1\times 10^{-5}}{\left(1-\mu^2\right)^2 \left(2+\mu^2\right)}\,.
\label{eq:thetaZpole}
\ee
The resulting constraint is shown in fig.\ref{fig:yukawa_constraints}.

\subsubsection{Search for sterile neutrinos in $4\ell$ final states at LEP-II}
\label{sec:wproduction}

\begin{figure}
\begin{center}
\includegraphics[scale=0.67]{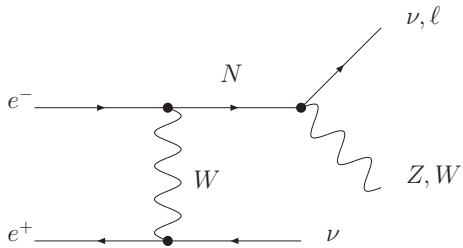}
\end{center}
\caption{Feynman diagram dominating the production of sterile neutrinos at the $WW$ threshold.}
\label{fig:nwthreshold}
\end{figure}

At LEP-II, the properties of $W$ bosons were studied at center of mass energies $\sqrt{s}$ at and beyond the $WW$ threshold. One of the relevant observables is the cross section for WW production, which can be reconstructed from the W decays into four-lepton final states,
\be
e^+e^- \to WW \to \bar \nu \ell^- \ell^+ \nu \,.
\label{eq:wproperties}
\ee
 The observed cross section was found to agree with the SM prediction and the Aleph experiment at LEP-II has placed a bound on possible SM deviations $\delta_{Aleph}$, defined via $|\delta \sigma_{WW\to 4\ell}| \leq \delta_{exp}\, \sigma_{WW\to 4\ell}^{\rm SM}$, at $1\sigma$ C.L.\ \cite{Heister:2004wr}: 
\be
\delta_{Aleph} = 1-\frac{n_{WW}^{Aleph}}{n_{WW}^{\rm SM}} = 0.005 \pm 0.011_{stat} \pm 0.007_{syst}\,.
\label{eq:Wboundthetae}
\ee
For our analysis we will combine the two contributing uncertainties in quadrature. For the SM prediction, we use the cross section for WW production from RacoonWW \cite{Denner:2000bj}. 

The dominant correction caused by sterile neutrinos to $e^+e^-\to 4\ell$ (at the considered energies) arises from diagrams of the type shown in fig.~\ref{fig:nwthreshold}, which produce the same final states from $N$ decays   
\be
e^+ e^- \to \bar \nu N \to \bar \nu \ell^- W^+ \to \bar \nu \ell^- \ell^+ \nu \,,
\ee
with a cross section $\sigma_{\nu N \to 4\ell}$. The produced four lepton final states would have been misinterpreted as as a contribution to WW pair production, which 
allows to constrain the sterile neutrino properties from the Aleph bound of eq.~(\ref{eq:Wboundthetae}). In the narrow width approximation, we obtain: 
\be
\sigma_{\nu N \to 4\ell} = \int \limits_{t_{\rm min}}^{t_{max}}dt \frac{d \sigma_{e^+e^-\to\nu N}}{dt} Br(N \to 3\ell)\,,
\ee
with $Br(N \to 3\ell) \simeq 0.2$ (not counting decays to neutrinos) and the kinematic limits $t_{min}=-s/2(1+\beta)+m_W^2$ and $t_{max}=-s/2(1-\beta)+m_W^2$. 
The differential cross section is given by \cite{Buchmuller:1991tu}
\bea
\frac{d \sigma_{e^+e^-\to\nu N}}{dt} & = \frac{G_F^2 \,m_W^4}{2\,\pi \, s^2} \sum\limits_{i=1}^3 & \left[2|\vartheta_{i\,4}|^2 t_W^4 \frac{1}{(s-m_Z^2)^2}\left(t(t-M^2)+u(u-M^2)\right) \right. \notag \\
 & & + \left|\frac{\vartheta_{i\,a}}{c_W^2} \frac{(1- 2\,s^2_W)}{2(s-m_Z^2)} - \frac{\theta_e {\cal N}^*_{e\,i}{\cal N}_{i \,e}^{}}{t-m_W^2}\right|^2 u\left(u-M^2\right) \notag \\
 & & \left.+ \left|\frac{\vartheta_{i\,a}}{c_W^2}\frac{(1- 2\,s^2_W)}{2(s-m_Z^2)} - \frac{\theta_e {\cal N}^*_{e\,i}{\cal N}_{i \,e}^{}}{u-m_W^2}\right|^2 t\left(t-M^2\right)\right]\,,
\label{eq:dsigma}
\eea
where $s_W,c_W,t_W$ are the sine, cosine and tangens of the weak mixing angle $\theta_W$, respectively. The bound on $|\theta_e|$ is then obtained from the requirement
\be
\sigma_{\nu N \to 4\ell} \leq  \delta_{exp}\, \sigma_{WW\to 4\ell}^{\rm SM}\,,
\ee
which is shown below in fig.~\ref{fig:wdecayimproved} by the black line and the grey area and in the summary plot fig.~\ref{fig:summarypresent}. 

A different analysis searching for sterile neutrinos beyond the $Z$ mass threshold has been conducted by the L3 collaboration \cite{Achard:2001qv}. They consider the production of active and sterile neutrino and the subsequent decay chain $N \to \ell_e\, W \to \ell_e \, j(j)$, where $j$ is a hadronic jet. The reconstructed invariant mass of the heavy neutrino would manifest as a peak in the invariant mass distribution. The bounds of \cite{Achard:2001qv} are of the same order as the ones presented here.

\subsubsection{Higgs boson decays at the LHC}

\begin{figure}
\begin{center}
\includegraphics[scale=0.5]{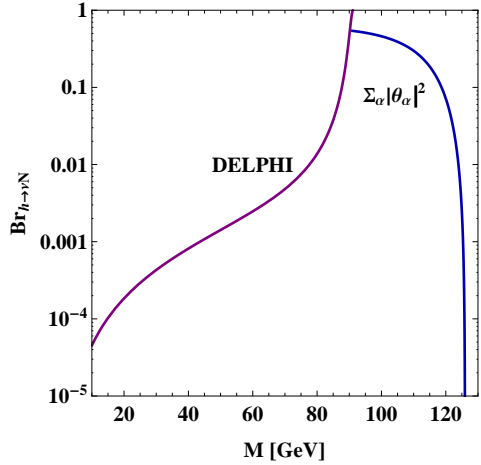}
\end{center}
\caption{Upper bound on the branching ratio of the Higgs boson into heavy and light neutrinos from ``indirect'' tests (solid blue line) and the direct search by Delphi at LEP-I \cite{Abreu:1996pa} (solid purple line). The branching ratio $Br_{h\to \nu N} := \sum_{i,j} Br(h\to  \nu_i N_j)$ denotes the sum of the processes $h\to \nu_i N_j$, for $i=1,2,3$ and $j=4,5$ and the Hermitian conjugate processes. For $M \gtrsim m_Z$ the present ``indirect'' constraints allow for even an order one $Br_{h\to \nu N}$.
}
\label{fig:yukawa}
\end{figure}

We now consider the constraints from the present LHC measurements of the Higgs decay parameters, which are shown in tab.~\ref{tab:LHChiggs}. Due to the large uncertainties on the fermionic branching ratios, we focus here on the decays $h\to VV$, for the vector bosons $V=\gamma,Z,W$. For $M \leq m_h$, the heavy neutrinos are produced in Higgs decays and will modify the branching ratios, which allows to constrain the neutrino Yukawa couplings (or equivalently the heavy-light mixing angles). There are basically two effects of the additional Higgs decays:

\begin{table}
\begin{center}
\begin{tabular}{|c|c|c|c|}
\hline
Channel &   $R_{\gamma\gamma}$	& $R_{WW}$	& $R_{ZZ}$  \\
\hline\hline
Atlas & {\small 1.17$^{+0.27}_{-0.27}$} & {\small 1.08$^{+0.22}_{-0.20}$} & {\small 1.44$^{+0.40}_{-0.33}$} \\
\hline
CMS & {\small 1.14$^{+0.30}_{-0.23}$} % 1407.0558
 & {\small 0.72$^{+0.20}_{-0.18}$} % 1312.1129
 & {\small 0.93$^{+0.29}_{-0.25}$} % 1312.5353
\\
\hline\hline
combined & 1.15(27) & 0.88(20) & 1.11(30) \\
\hline
\end{tabular}
\end{center}
\caption{Currently best measured decay ratios $R_{XX} = Br(h\to XX)^{\rm exp}/Br(h\to XX)^{\rm SM}$ from CMS \cite{Khachatryan:2014ira,Chatrchyan:2013iaa,Chatrchyan:2013mxa} and ATLAS \cite{Aad:2013wqa}. 
}
\label{tab:LHChiggs}
\end{table}

Firstly, the total Higgs decay width is enlarged, which effectively reduces all the SM branching ratios by a factor
\be
r = \frac{\Gamma_{h,\,\rm SM}}{\Gamma_{h,\,\rm SM} +\Gamma_{h\to \nu N}}\,.
\ee
In fig.~\ref{fig:yukawa} the upper bound on the branching ratio for the total decay width of the Higgs boson into light and a heavy neutrinos is shown by the solid blue line.  The plot shows that even an {\cal O}(1) branching ratio $Br_{h\to \nu N}$ would be consistent with the present ``indirect'' constraints. 

Secondly, when heavy neutrino decays take place inside the detector (which is the case for the relevant parameter space we are considering here), then the subsequent decays may get counted as SM Higgs decays into $ZZ$ or $WW$ since they lead to the same final states. Altogether, the experimentally measured branching ratios with respect to the SM prediction are given by
\be
Br_{h\to XX} = r\, Br_{h\to XX,\,\rm SM} + c_X Br_{h\to \nu N}, \qquad \text{with} \qquad c_X = \left\{\begin{array}{ll} \frac{1}{2}, & X=Z,W \\ 0, & X=\gamma,\, f
\end{array}
\right.\,,
\label{eq:higgsbranchingdev}
\ee
with $Br_{h\to \nu N} = \Gamma_{h\to \nu N}/(\Gamma_{h,\,\rm SM} + \Gamma_{h\to \nu N})$.
The experimental precision of the currently best measured observables (cf.\ tab.~\ref{tab:LHChiggs}) can now be translated into sensitivities on the active-sterile mixing parameters. 

\begin{figure}
\begin{center}
\includegraphics[scale=0.6]{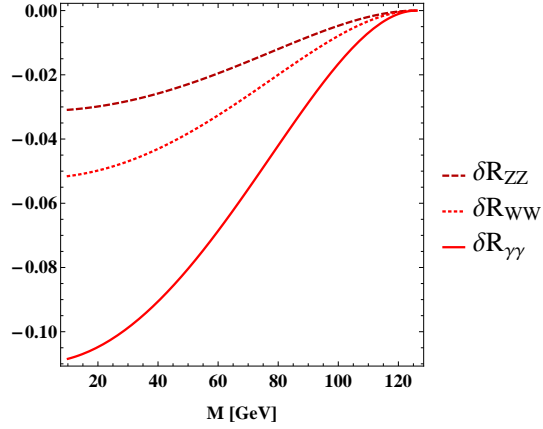}
\end{center}
\caption{Deviation of the Higgs decay ratios $R_{XX}$ from the SM prediction $R_{XX} = 1$ for the example value $\sum_\alpha |y_{\nu_\alpha}|^2 = 10^{-4}$. }
\label{fig:Higgsdeviation}
\end{figure}

\begin{figure}
\begin{minipage}{0.49\textwidth}
\begin{center}
\includegraphics[scale=0.63]{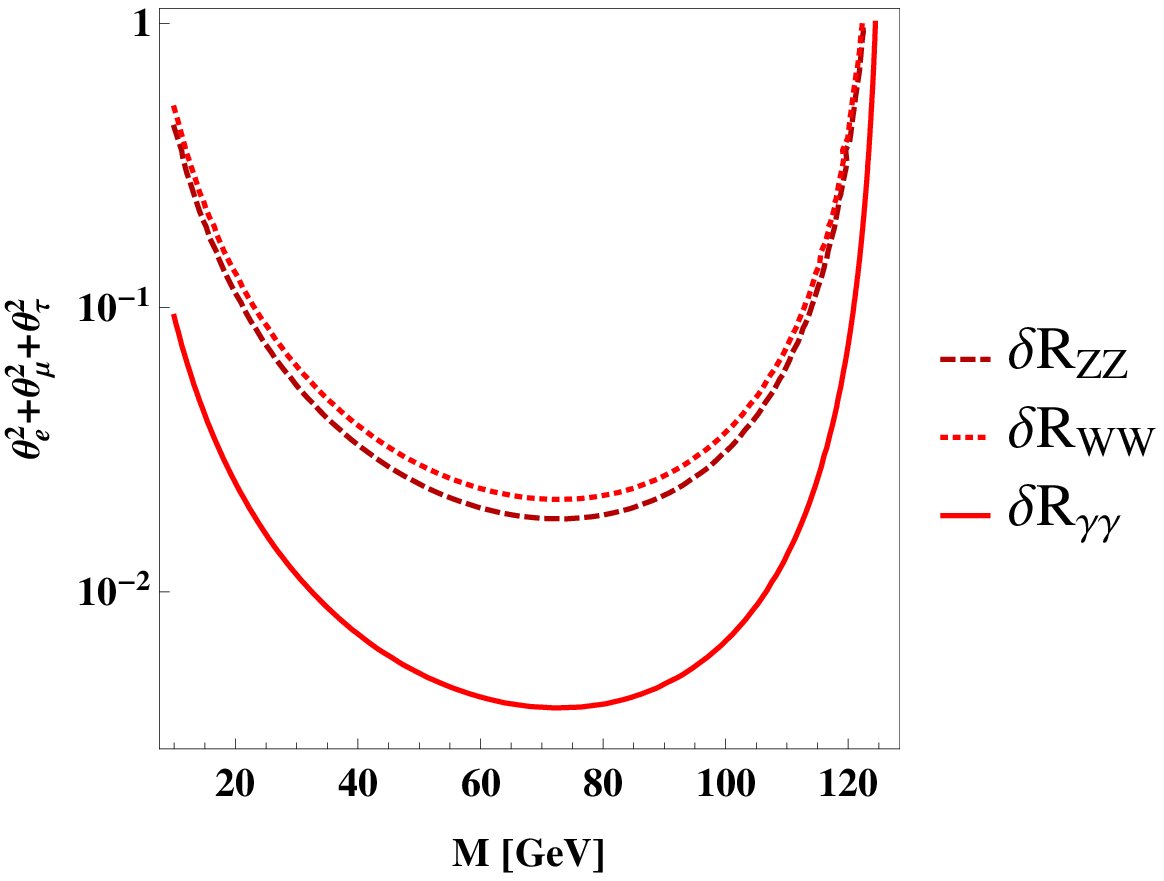}
\end{center}
\end{minipage}
\begin{minipage}{0.49\textwidth}
\begin{center}
\includegraphics[scale=0.45]{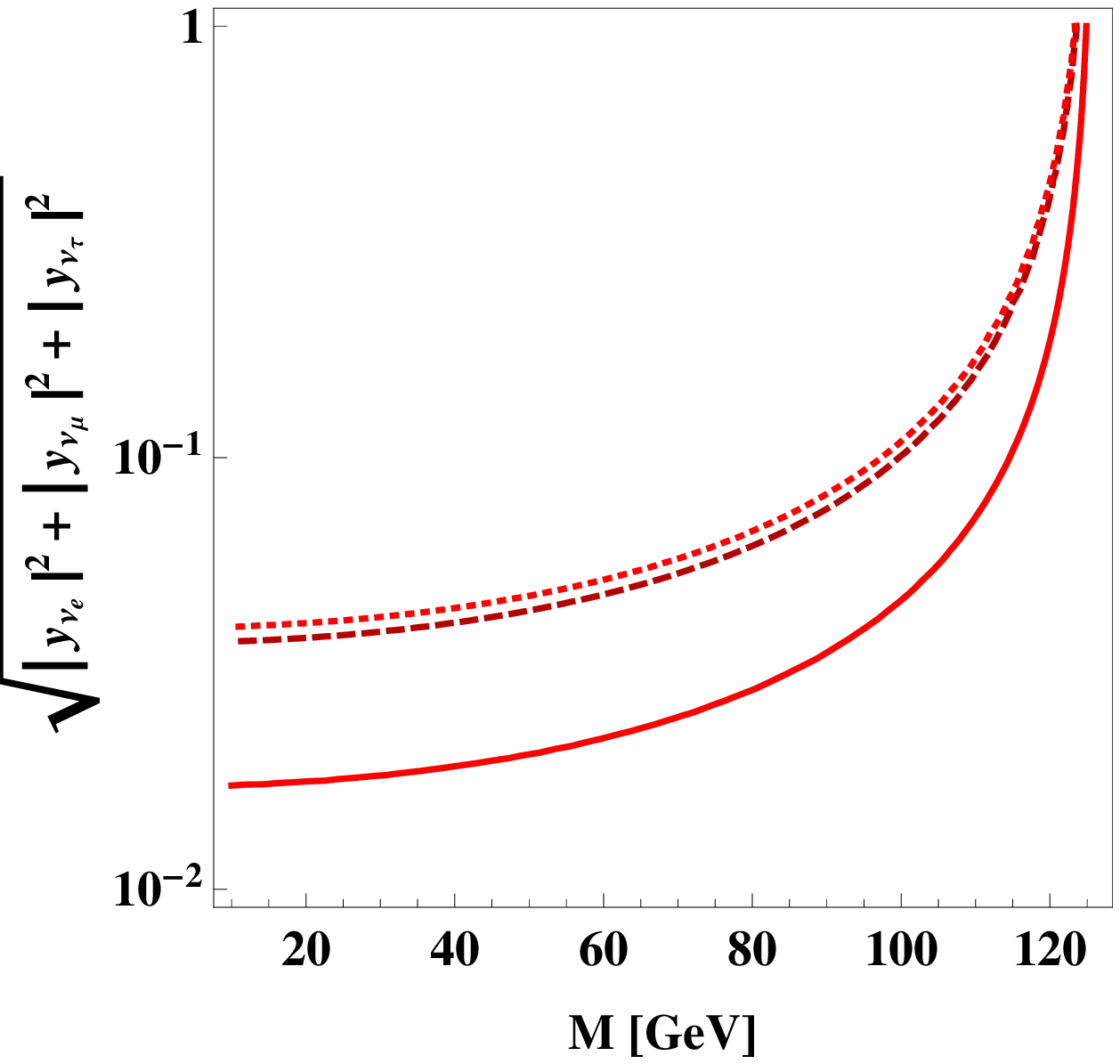}
\end{center}
\end{minipage}

\caption{
Constraints on sterile neutrino parameters from Higgs decays at the LHC.
}
\label{fig:branching}
\end{figure}

In fig.~\ref{fig:Higgsdeviation} we display the deviation of the Higgs branching ratios into gauge bosons, for the squared sum of Yukawa couplings $\sum_\alpha |y_{\nu_\alpha}|^2=10^{-4}$, for illustration. Figure \ref{fig:branching} shows the sensitivity of the present LHC measurements on the active-sterile mixing parameters for the different decay channels, the most sensitive of which is $Br(h\to \gamma\gamma)$. Note that when estimating the constraints we assumed that every single heavy neutrino decay is counted into the associated decay channel of Higgs to a gauge boson. In reality, some events might be removed by the experimental filters. The assumption we made leads to a conservative bound, since the two effects described above change the branching ratios in opposite directions. Successful filtering would increase the sensitivity of the branching ratios into $W$ and $Z$. The estimates obtained here are in good qualitative agreement with the analyses in refs.~\cite{Cely:2012bz} and \cite{BhupalDev:2012zg}, where the event signature has been analysed more carefully.

Other sterile neutrino decay signatures at the LHC have been analysed e.g.\ in refs.\cite{delAguila:2008hw,delAguila:2009bb,Helo:2013esa,Das:2014jxa,Das:2012ze}.

\section{Possible improvements from future lepton colliders}

We will now estimate how the improved sensitivities of future colliders could allow to test sterile neutrino properties. The improvements will concern ``indirect'' tests, especially via the EWPOs, as well as ``direct'' searches. We will focus on the processes discussed in the previous section and consider the ILC, CEPC and FCC-ee (TLEP), which are currently discussed, as representative examples. We note already here that the numbers which are given are based on estimates for sensitivities as they are currently discussed in the respective working groups or are extracted from present proposals. These numbers may change and affect the comparison between the experiments. 

For instance, ref.~\cite{Fan:2014vta} has recently suggested for the CEPC, to increase the integrated luminosity off the $Z$ peak, and the use of polarised beams, in order to boost the precision of the EWPO measurements. An important result of this suggestion is the improvemed precision when measuring $s_{W,eff}$, which we include in the third column of tab.\ref{tab:flcprec}.

The table also includes the current estimates for the possible experimental precision supplied by the CEPC study group\footnote{We thank M. Ruan at this point for support with CEPC machine parameters.}.  We remark however, that not all information regarding the CEPC performance in the electroweak precision sector are available at present. For a comparative study of the machine performance we include a second column on the CEPC uncertainties in tab.~\ref{tab:flcprec}, which contains the precision of the FCC-ee, scaled with a factor of $\sqrt{10}$. This would correspond to a CEPC with identical performance parameters as the FCC-ee but with $10^{11}$ $Z$ bosons instead of $10^{12}$ as considered in \cite{Gomez-Ceballos:2013zzn} for the FCC-ee.

\subsection{Sensitivities of ``indirect'' searches at future colliders: EWPOs}

\begin{figure}
\begin{minipage}{0.49\textwidth}
\begin{center}
\includegraphics[scale=0.6]{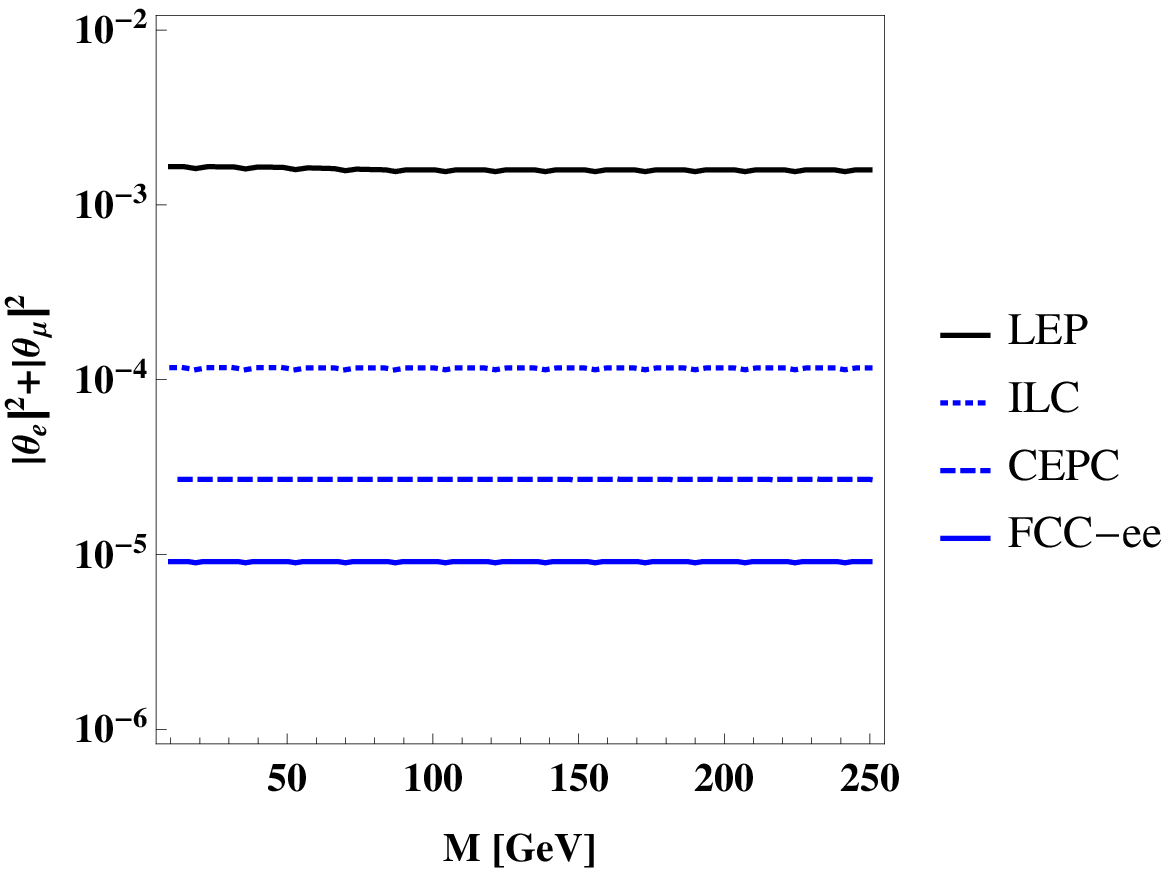}
\end{center}
\end{minipage}
\begin{minipage}{0.49\textwidth}
\begin{center}
\includegraphics[scale=0.45]{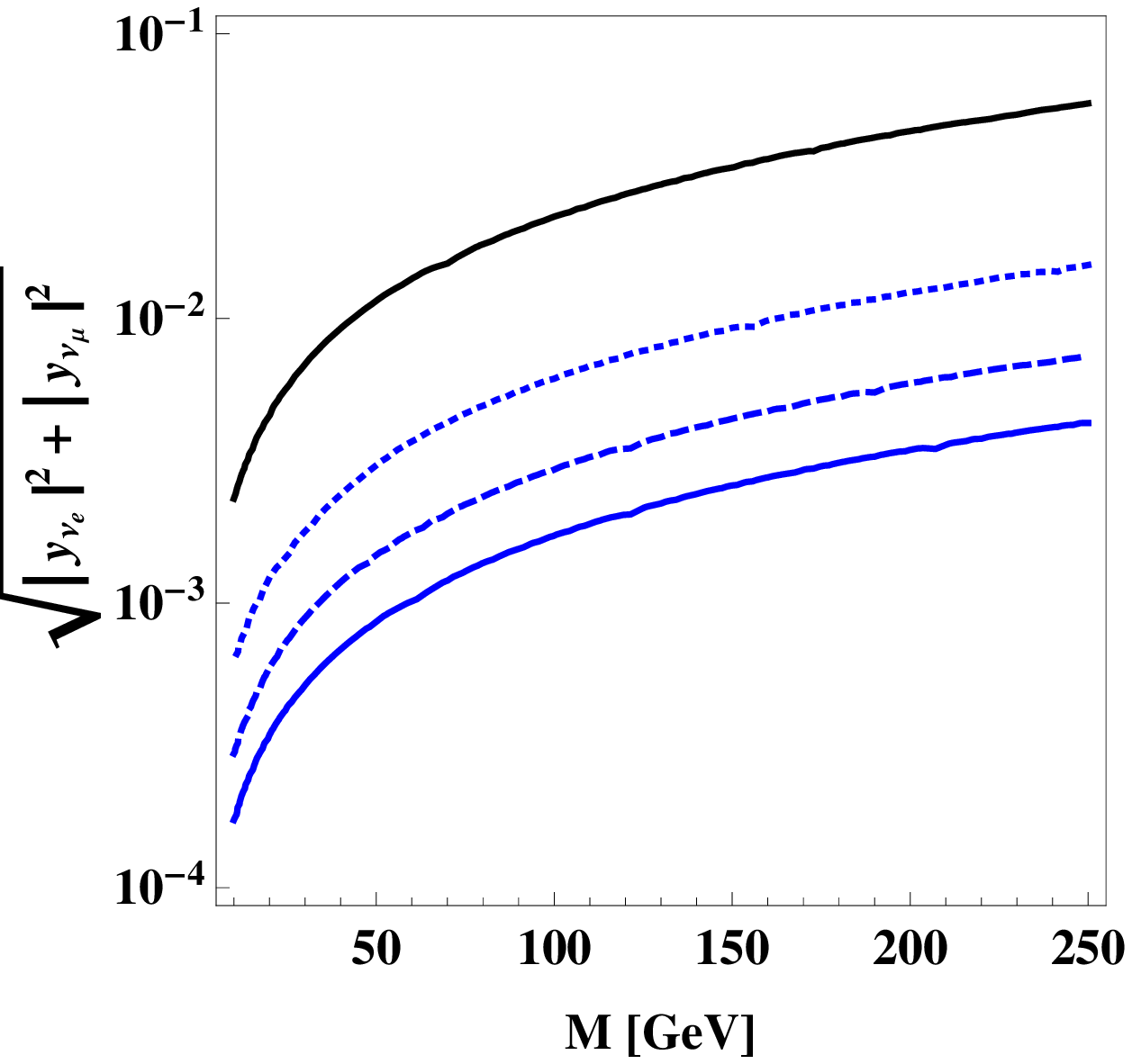}
\end{center}
\end{minipage}

\begin{minipage}{0.49\textwidth}
\begin{center}
\includegraphics[scale=0.6]{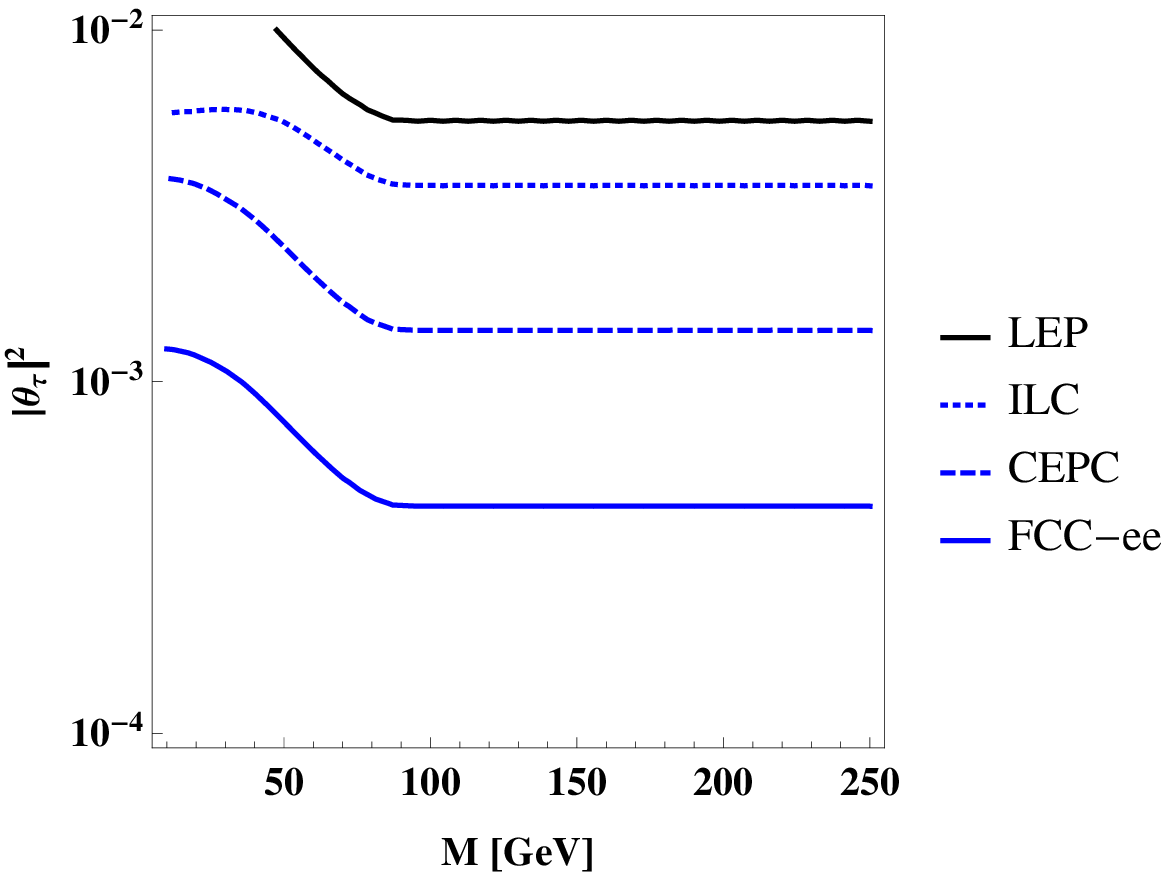}
\end{center}
\end{minipage}
\begin{minipage}{0.49\textwidth}
\begin{center}
\includegraphics[scale=0.45]{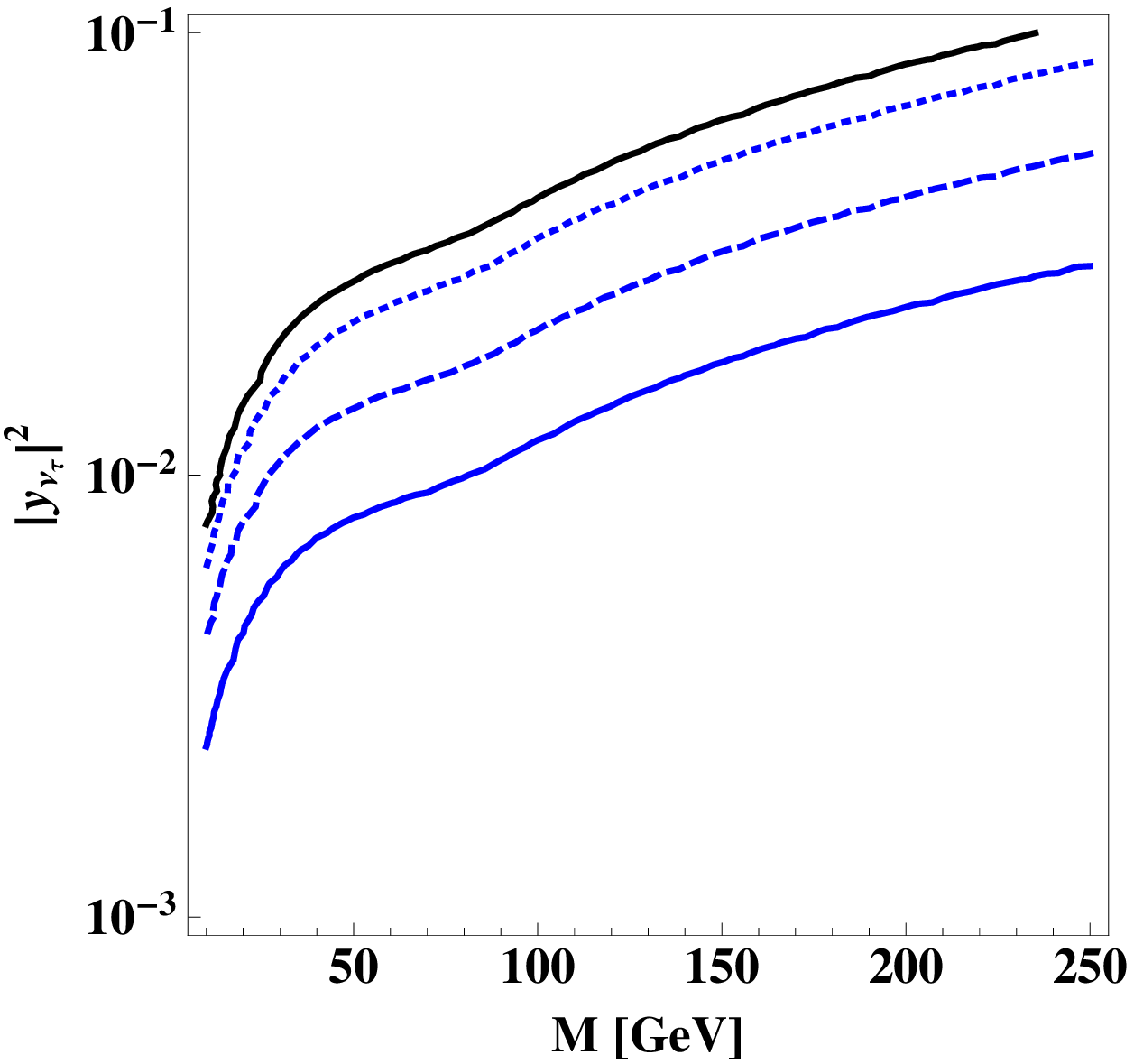}
\end{center}
\end{minipage}
\caption{Estimated sensitivities on the active-sterile mixing parameters at 90\% confidence level from the EWPOs, using the uncertainties given in tab.~\ref{tab:flcprec}. For the CEPC, we use the fourth column labeled CEPC$^*$.
}
\label{fig:boundyukawa}
\end{figure}

As discussed in the previous section, the EWPOs are sensitive to the parameter combinations $|\theta_\tau|^2$ and $|\theta_e|^2+|\theta_\mu|^2$ for a given $M$. For estimating the possible future sensitvities we use the observables and experimental uncertainties listed in tab.~\ref{tab:flcprec}. 
The resulting bounds on the parameters $|\theta_\tau|^2$ and $|\theta_e|^2+|\theta_\mu|^2$ and the bounds on the Yukawa couplings $|y_{\nu_\tau}|$ and $\sqrt{|y_{\nu_e}|^2+|y_{\nu_\mu}|^2}$ are shown in the four plots of fig.~\ref{fig:boundyukawa}. Note the relaxed constraint on $|\theta_\tau|$ (and $|y_{\nu_\tau}|$) for $M < m_Z$, due to the phase-space factors in the theoretical prediction for the EWPOs.

\begin{table}
\begin{center}
\begin{tabular}{|l|c|c|c|c|}
\hline
Observable & ILC & FCC-ee & CEPC & CEPC$^*$\\
\hline\hline
$R_\ell$ & 0.004 & 0.001 & 0.01 & 0.003$^*$ \\
$R_{inv}$ & 0.01 & 0.002 & 0.012 & 0.006$^*$\\
$R_b$ & 0.0002 & 0.00002 & 0.00017 & 0.0007$^*$\\
$M_W$ [MeV]    & 2.5   & 0.5 & 0.5 & 0.5 \\
$s_{eff}^{2,\ell}$  & 1.3 $\times 10^{-5}$  & 1 $\times 10^{-6}$ & 2.3 $\times 10^{-5}$ & 3.3 $\times 10^{-6}{}^*$ \\
$\sigma_h^0$ [nb]  &  0.025  &  0.0025 & -- & 0.008$^*$ \\
$\Gamma_\ell$ [MeV]   & 0.042  & 0.0042 & -- & 0.014$^*$ \\
\hline\hline
Reference & \cite{Baak:2013fwa} & \cite{Gomez-Ceballos:2013zzn} & \cite{Ruan:2014xxa}, \cite{Fan:2014vta} & scaled$^*$ \\
\hline
\end{tabular}
\end{center}
\caption{Estimated systematic uncertainties of the ILC, the CEPC and the FCC-ee for future measurements of the EWPOs. (The statistical uncertainties would be much smaller.) \\
$^{*)}$ Performance scaled with a factor $\sqrt{10}$ from FCC-ee, for comparison. See text for details.}
\label{tab:flcprec}
\end{table}

\subsection{Sensitivities of  ``direct'' tests at future colliders}

\subsubsection{Future searches for sterile neutrinos produced in $Z$ boson decays}

\begin{figure}
\begin{minipage}{0.6\textwidth}
\begin{center}
\includegraphics[scale=0.6]{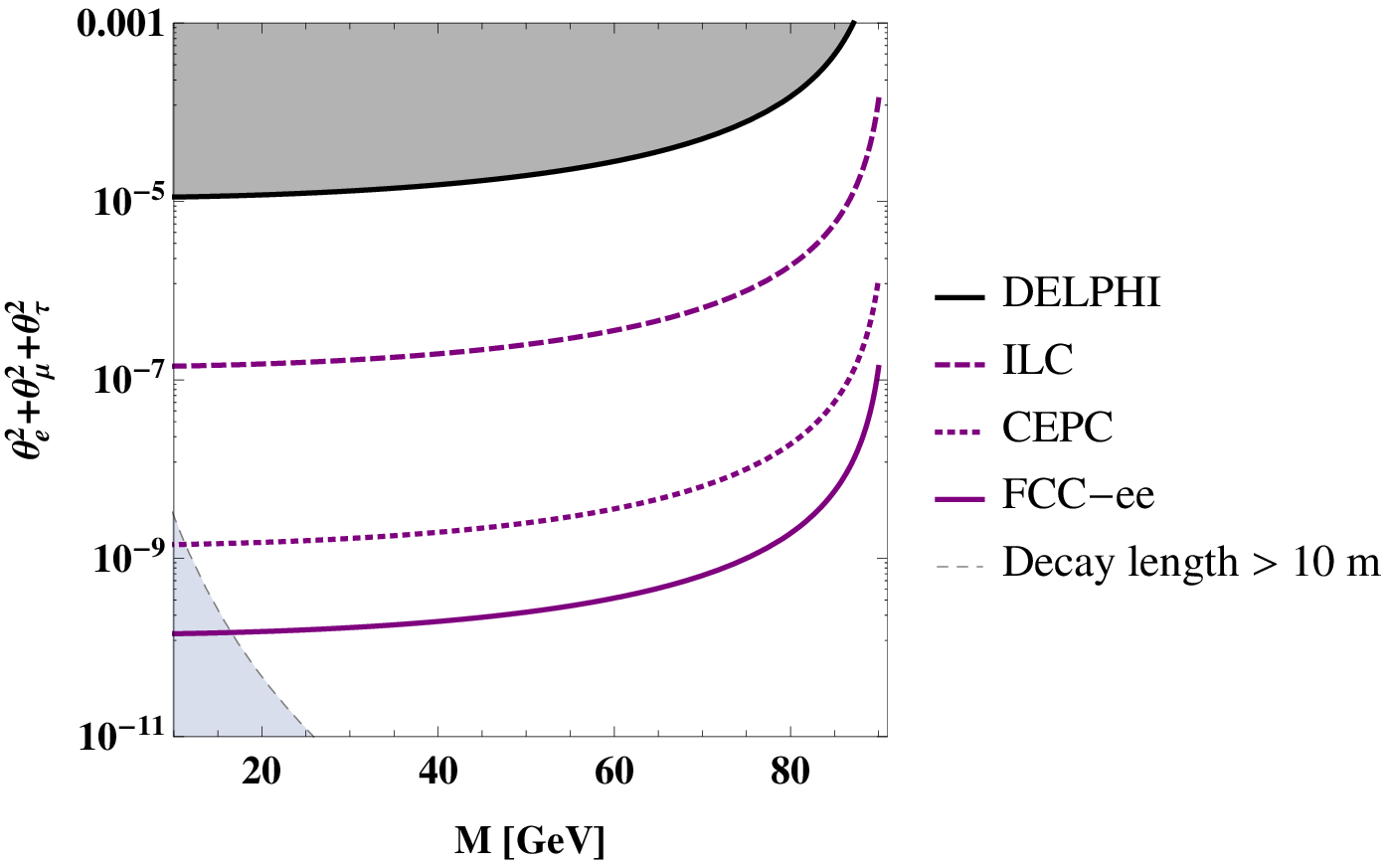}
\end{center}
\end{minipage}
\begin{minipage}{0.39\textwidth}
\begin{center}
\includegraphics[scale=0.45]{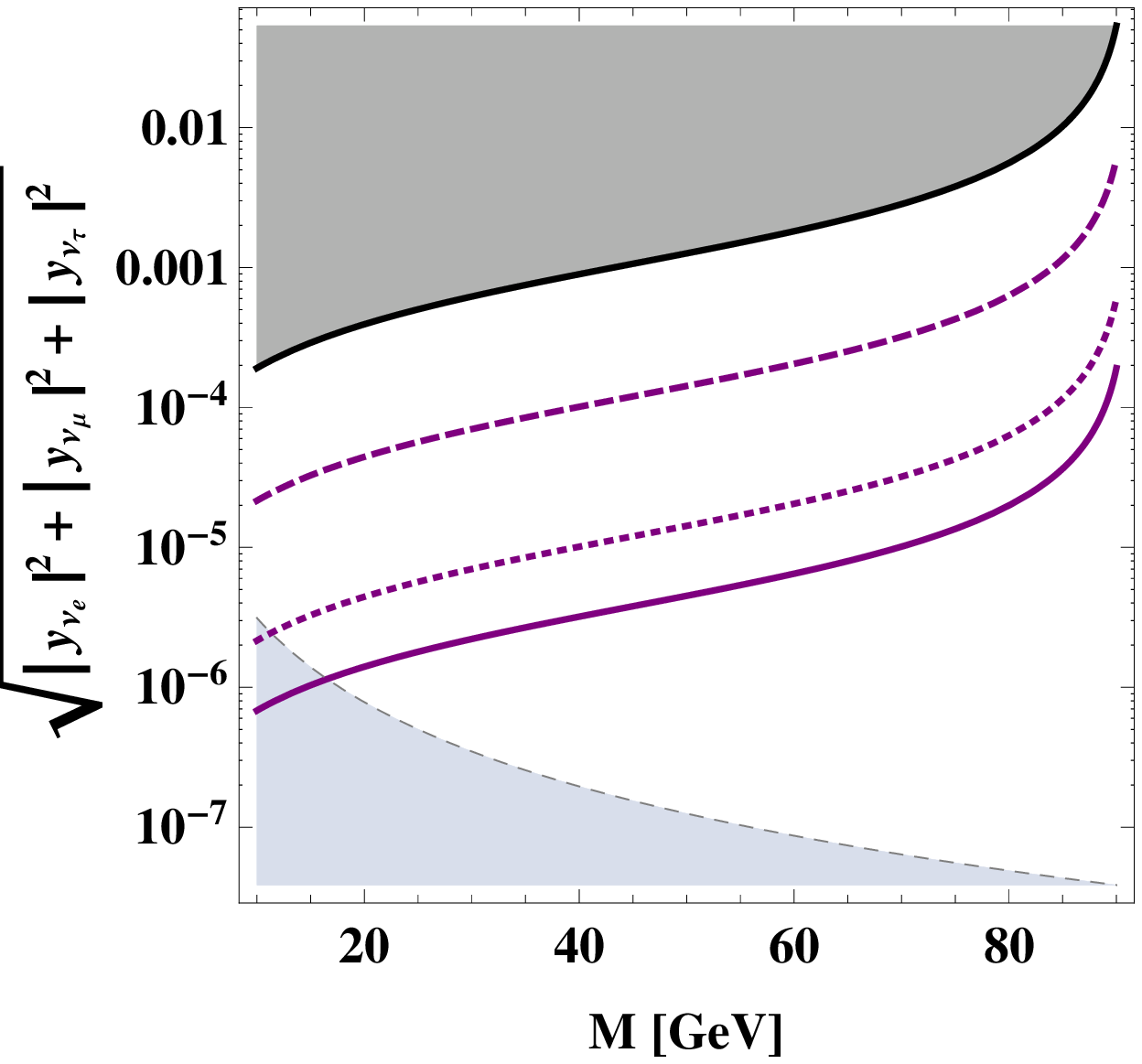}
\end{center}
\end{minipage}
\caption{Estimated sensitivities at 95\% confidence level of a search for heavy neutral leptons at the $Z$ pole, analogous to the one performed at LEP-I by Delphi \cite{Abreu:1996pa}. For model parameters in the lighter grey area, the heavy neutrinos may decay more than 10 meters away from the primary vertex and thus become invisible to the detector (cf.\  \cite{Gronau:1984ct}).}
\label{fig:Zsearch}
\end{figure}

The estimated number of $Z$ bosons produced by the future lepton colliders are $10^{9}$ at the ILC \cite{Baak:2013fwa} and $10^{11}$ at CEPC \cite{Ruan:2014xxa}. 
For the FCC-ee we use $10^{13}$ produced $Z$ bosons as discussed in \cite{Blondel:2014bra}. In analogy to the Delphi analysis discussed in sec.~\ref{sec:delphi}, we estimate the bound on $\theta^2$ achievable with a larger sample of $Z$ bosons by scaling eq.~(\ref{eq:thetaZpole}), if no deviation from the SM should be found: 
\be
\theta^2 \leq \frac{1.4}{\left(1-\mu_Z^2\right)^2 \left(2+\mu_Z^2\right)} \times  \left\{ \begin{array}{cl} 10^{-11} & \text{FCC-ee} \\ 10^{-9} & \text{CEPC} \\ 10^{-7} & \text{ILC} \end{array}\right. \,,
\ee
where $\mu_Z = M/m_Z$. We show the resulting sensitivity for the future lepton colliders, based on this estimate, in fig.~\ref{fig:Zsearch}. The region in parameter space, where the decays of the sterile neutrinos take place more than ten meters away from the primary vertex, are shown by the grey area in the figure. We used the formula from ref.~\cite{Gronau:1984ct}. We note that even stronger bounds could be possible for $M$ close to the grey shaded region, from searches for displaced vertices, as discussed for the FCC-ee in \cite{Blondel:2014bra}.

\subsubsection{Searches for sterile neutrinos in $4\ell$ final states at 250 GeV}
One of the central aspects of a future lepton collider is the precise measurement of the Higgs boson properties. The dominant Higgs production mechanism is Higgs-strahlung,  at center of mass energies around 250 GeV, which implies that large quantities of $W$ pairs will be produced as a byproduct.
Analogous to the analysis of the LEP-II data in sec.~\ref{sec:wproduction}, a future measurement of the cross section of the process $e^+e^- \to 4 \ell$  can be used to constrain sterile neutrino properties. 

\begin{figure}
\begin{minipage}{0.49\textwidth}
\begin{center}
\includegraphics[scale=0.6]{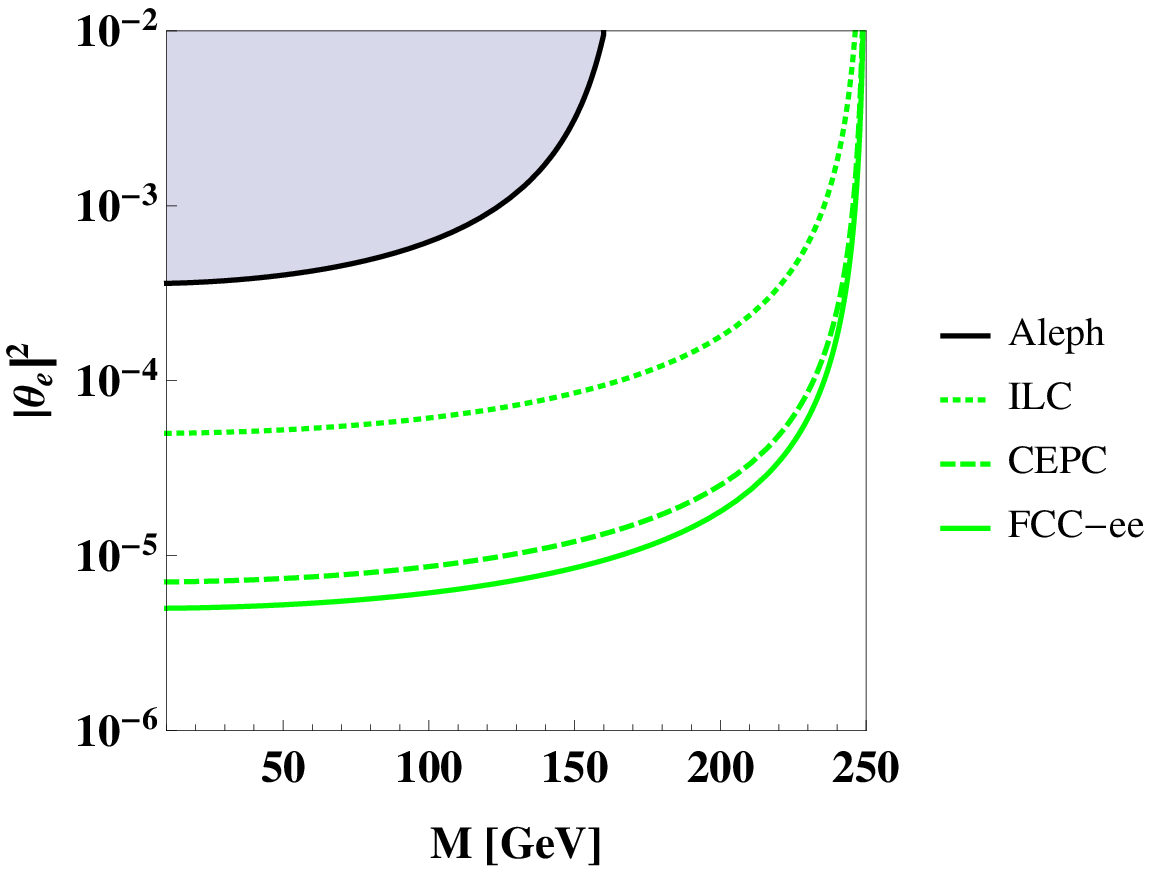}
\end{center}
\end{minipage}
\begin{minipage}{0.49\textwidth}
\begin{center}
\includegraphics[scale=0.45]{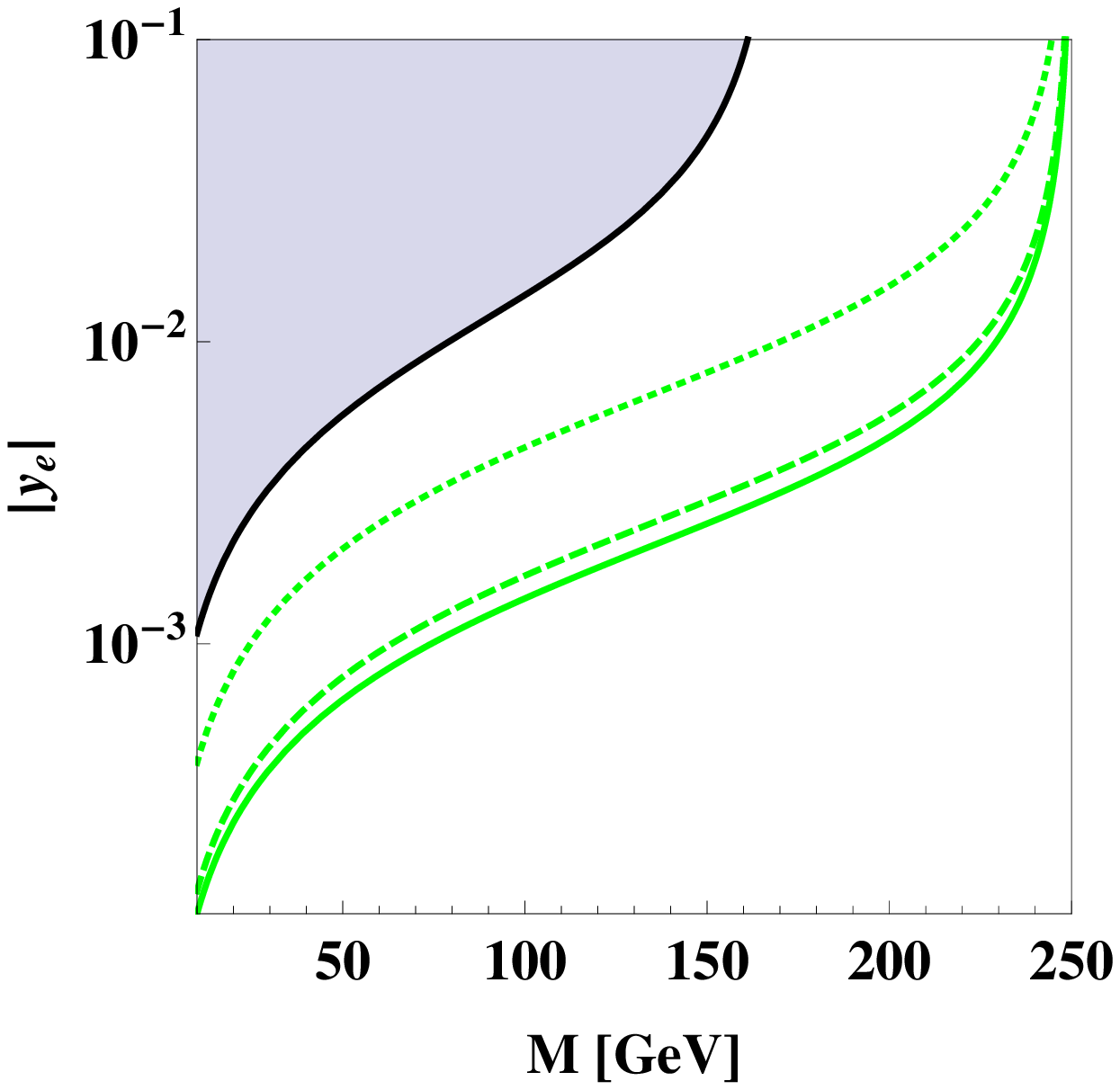}
\end{center}
\end{minipage}
\caption{Estimated sensitivity on the sterile neutrino parameters from $e^+ e^- \to 4 \ell$ at the $WW$ threshold and beyond. The black line shows the Aleph constraint for comparison.}
\label{fig:wdecayimproved}
\end{figure}

The estimated $W$ boson yield of the ILC, CEPC and FCC-ee is shown in tab.~\ref{tab:Wdecayprecision}, where we include Aleph for comparison. At present, no official estimates for the systematic uncertainties are available, but discussions in the working groups are ongoing. We therefore only consider the statistical uncertainty for calculating our estimate. We stress that the estimated constraints from $W$ boson measurements should therefore be taken with caution. They rather correspond to a maximally reachable sensitivity. The estimates for the constraints will be updated as soon as official forecasts for the systematic uncertainties are available. Given this warning, we present the estimated sensitivities on the active-sterile neutrino mixing parameter $|\theta_e|$ (and $|y_{\nu_e}|$) from eq.~(\ref{eq:Wboundthetae}) in fig.~\ref{fig:wdecayimproved}.

\begin{table}
\begin{center}
\begin{tabular}{|c|c|c|c|c|}
\hline
 & Aleph & ILC & CEPC & FCC-ee \\
\hline\hline
\#$W$'s prod. & $10^4$ & $10^7$ & $10^8$ & $2\times 10^8$ \\
\hline
$\delta_{\rm stat.}$ on $\sigma_{WW\to 4\ell}^{\rm SM}$ & $10^{-2}$ & 3$\times 10^{-4}$ & $10^{-4}$ & $7\times10^{-5}$\\
\hline 
\end{tabular}
\end{center}
\caption{Expected number of $W$ bosons produced at the considered future colliders and at Aleph for comparison. We assume that the production takes place at $\sqrt{s}=250$ GeV, apart from Aleph, where $\sqrt{s}=$ 161 to 209 GeV. $\delta_{\rm stat.}$ denotes the statistical uncertainty on the measurement of $\sigma_{WW\to 4\ell}^{\rm SM}$, which we use for our estimates.}
\label{tab:Wdecayprecision}
\end{table}

\subsubsection{Sensitivities of future measurements of Higgs boson branching ratios} 
The sensitivity to the Higgs boson properties at 240--250 GeV can be significantly improved at a future lepton collider, compared to the LHC. The currently available estimated precision of the ILC, CEPC and FCC-ee are shown in the tab.~\ref{tab:higgsprecision}. 

\begin{figure}
\begin{minipage}{0.49\textwidth}
\begin{center}
\includegraphics[scale=0.65]{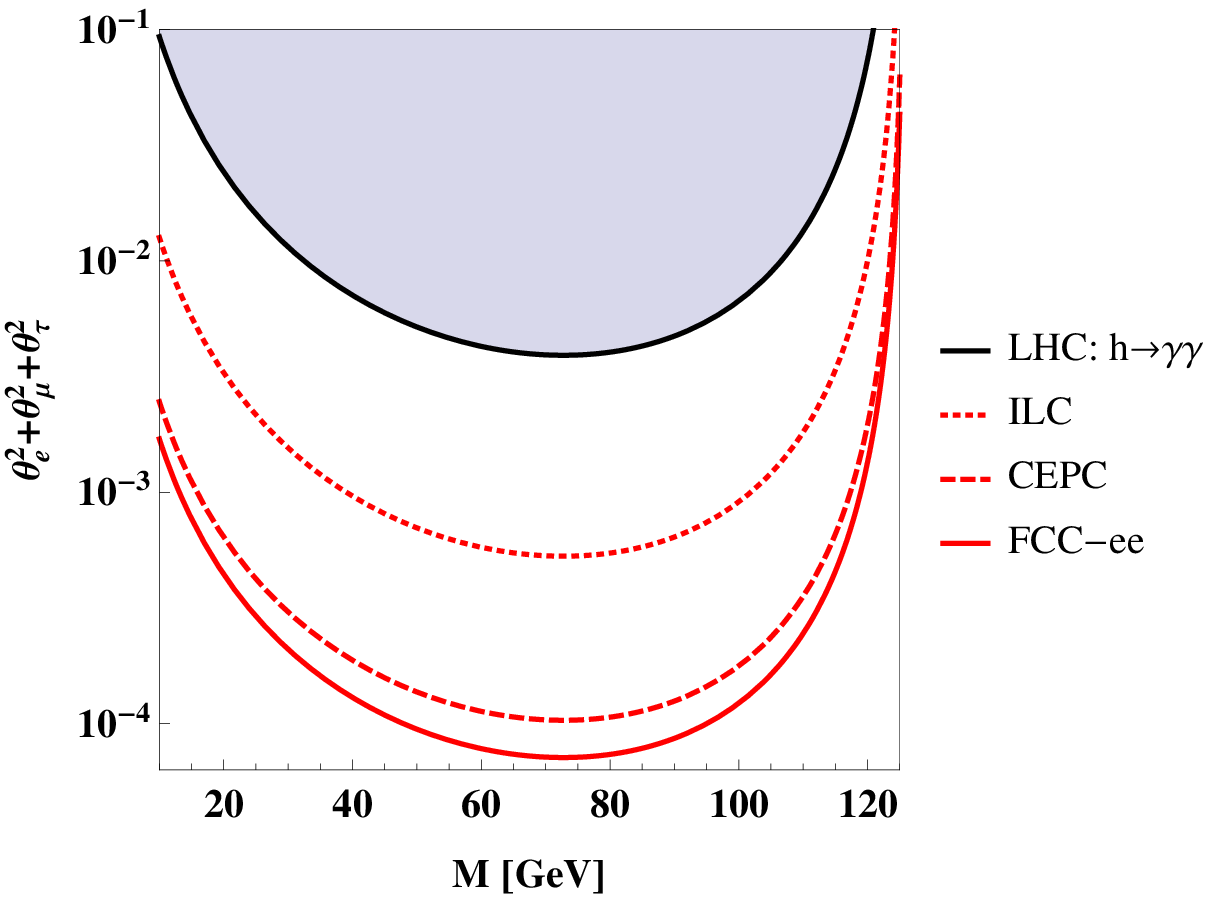}
\end{center}
\end{minipage}
\begin{minipage}{0.49\textwidth}
\begin{center}
\includegraphics[scale=0.475]{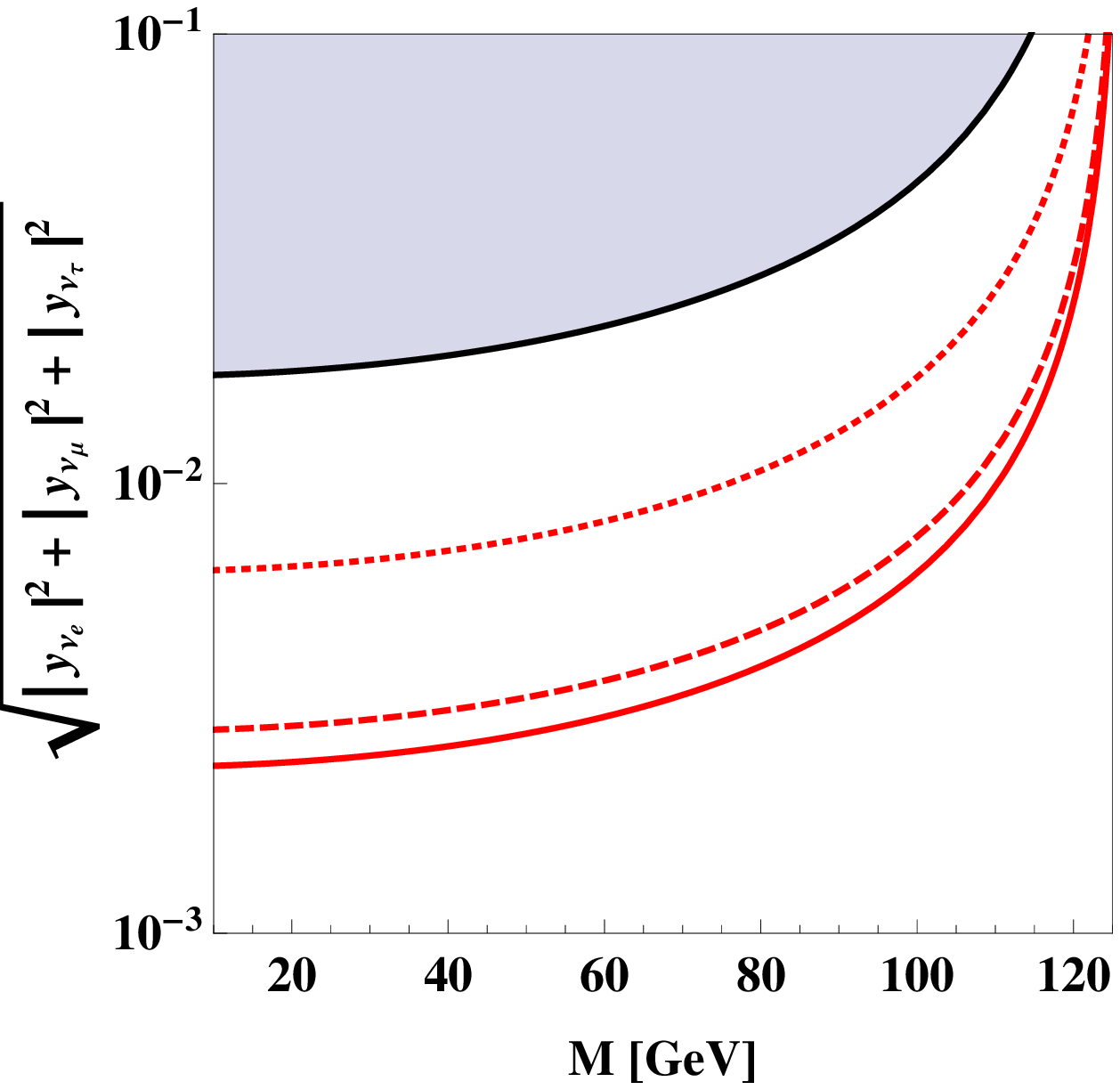}
\end{center}
\end{minipage}
\caption{Estimated sensitivities on the sterile neutrino properties from the decays of the Higgs boson to $W$ bosons, which is the Higgs decay channel most sensitive to heavy neutrinos at future lepton colliders, assuming 10 years of data taking. The black line denotes the present bounds from the LHC coming from $h\to \gamma \gamma$.}
\label{fig:Higgssearch}
\end{figure}

For estimating the future sensitivity to sterile neutrino properties, we consider the Higgs branching with the highest precision, namely $Br_{h\to WW}$, in tab.~\ref{tab:higgsprecision}. The results are shown in fig.~\ref{fig:Higgssearch}.  A comparison of tab.~\ref{tab:higgsprecision} with fig.~\ref{fig:yukawa}, which includes the currently allowed branching ratios into heavy and light neutrinos, shows that with the increased precision the future measurements are indeed sensitive to deviations caused by sterile neutrinos, despite the already strong constraints from precision data and from the Delphi experiment.

Furthermore, the branching ratio $Br_{h\to \mathrm{invisible}}$ also provides a promising channel for sterile neutrino searches, if the sensitivity for the branching ratio could reach $0.1$\%. In this case it would be complementary and comparable in sensitivity to $Br_{h\to WW}$.

\subsubsection{Sensitivities of future measurements of $\boldsymbol{e^+e^- \to h + \slashed{E}_T}$}

\begin{figure}
\begin{minipage}{0.49\textwidth}
\begin{center}
\includegraphics[scale=0.47]{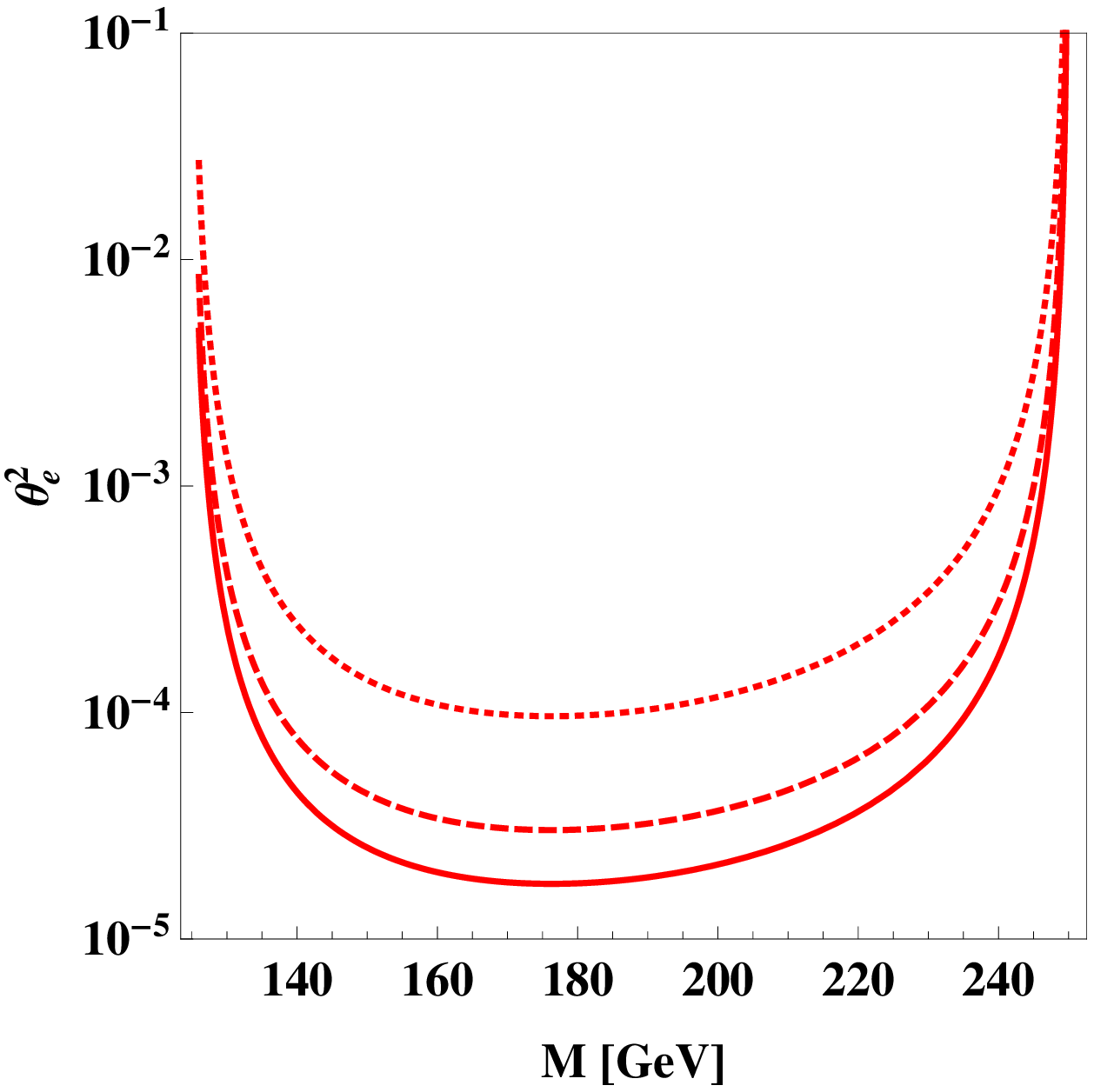}
\end{center}
\end{minipage}
\begin{minipage}{0.49\textwidth}
\begin{center}
\includegraphics[scale=0.49]{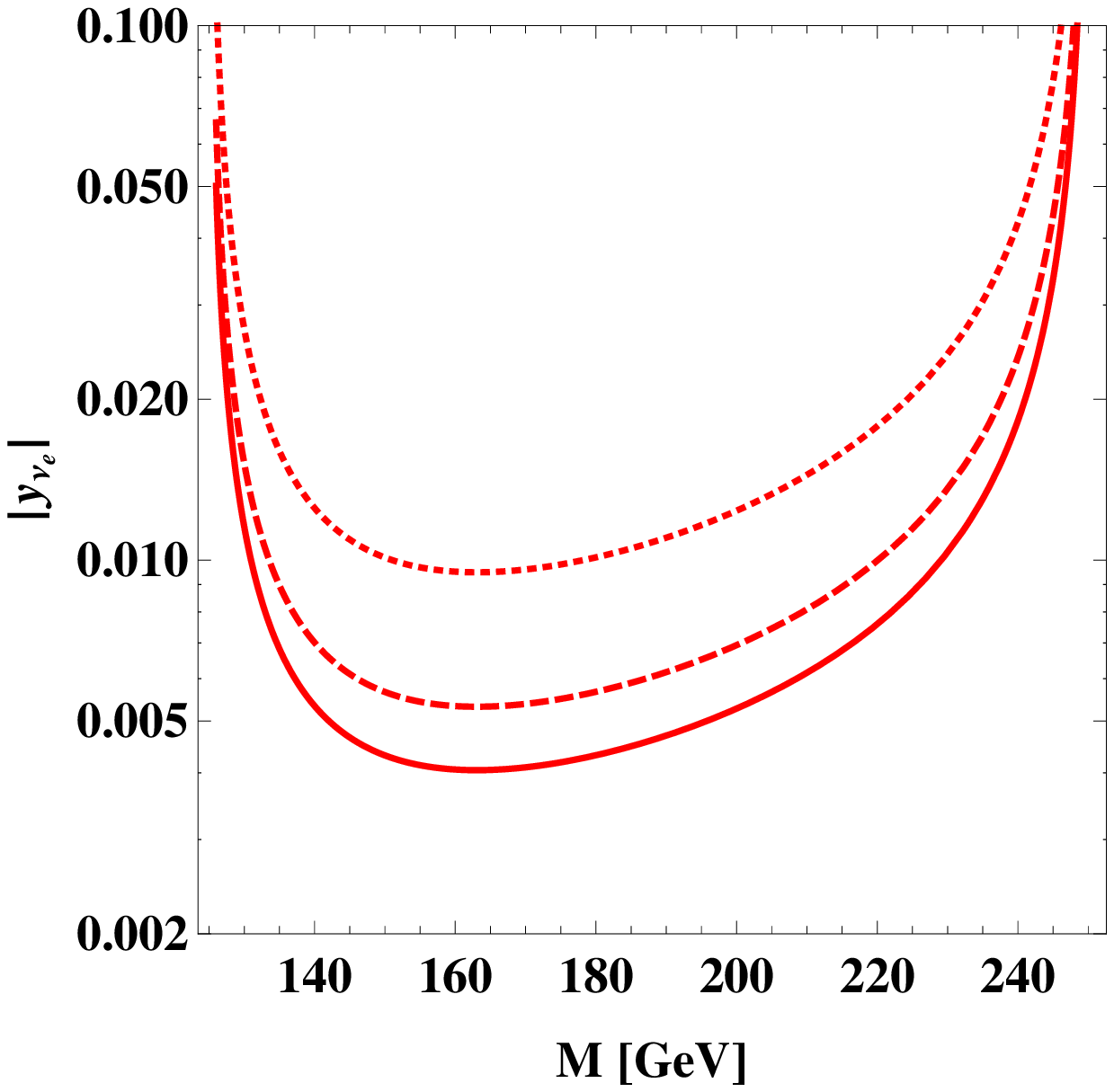}
\end{center}
\end{minipage}
\caption{Estimated sensitivities on the sterile neutrino properties from $e^+e^- \to h + \slashed{E}_T$. }
\label{fig:Higgssearch2}
\end{figure}

Another sensitive channel to search for sterile neutrino signals is the process electron-positron to Higgs boson plus missing transverse energy.\footnote{We would like to thank W. Murray making us aware of this possibility.} The SM background is given by $e^+e^- \to Z^* \to Z\, h$ with the subsequent decay $Z \to \bar \nu \nu$. The Higgs boson is tagged via two $b$-jets. In the presence of sterile neutrinos, a light and a heavy neutrino can be produced by $Z$ and $W$ exchange as discussed above. The sterile neutrino then decays into a light neutrino and a Higgs boson, thus contributing to $Br_{e^+ e^- \to h+ \slashed{E}_T}$. This process is relevant for $M > m_H$. The estimated future experimental precision for measuring this branching ratio is included in tab.~\ref{tab:higgsprecision} and the estimated sensitivities for the sterile neutrino properties are displayed in fig.~\ref{fig:Higgssearch2}.

\begin{table}
\begin{center}
\begin{tabular}{|c||c|c|c|}
\hline
Branching ratio & ILC & CEPC & FCC-ee \\
\hline\hline
$Br_{h\to WW}$	& 6.4 & 1.3 & 0.9 \\
\hline
$Br_{h\to ZZ}$	& 19 & 5.1 & 3.1 \\
\hline
$Br_{h\to \gamma\gamma}$ & 35 & 8 & 3.0 \\
\hline
$Br_{e^+ e^- \to h+ \slashed{E}_T}$ & 11.0$^*$	& 3.8 	& 2.2 \\
\hline
\end{tabular}
\end{center}
\caption{Estimated precision for the measurement of the Higgs boson branching ratios at future lepton colliders, for one year of running. The numbers are in percent, and taken from refs.~\cite{Baak:2013fwa,Gomez-Ceballos:2013zzn,Ruan:2014xxa}. \\ $^{*)}$ Estimated value obtained from the FCC-ee estimate rescaled with the ILC luminosity.
}
\label{tab:higgsprecision}
\end{table}

\subsection{Expected future sensitivity to sterile neutrino properties from charged lepton flavour violation and neutrino oscillation experiments}

The current bound on the charged lepton flavour violating decay $\mu \to e \gamma$ from the MEG experiment \cite{Adam:2013mnn} plays an important role in the present ``indirect'' probes of sterile neutrino properties, which we presented as results of a global fit in section \ref{sec:indirectconstraints}. The MEG bound constrains the product $|\theta_e\theta_\mu| <  1.5 \times10^{-5}/\sqrt{1-0.3 F(x_M)}$, with $F(x_M)$ defined in eq.~(\ref{eq:loopfunction}). In the global fit, with a non-zero best fit value for $| \theta_e |$, it drives the tight constraint on $| \theta_\mu |$.

Future tests of the $\mu \to 3e$ branching ratio from Mu3e \cite{Blondel:2013ia} and MUSIC \cite{Ogitsu:2011rg,Yamamoto:2011zb} and tests of atomic conversion rate of $\mu \to e$ from Mu2e \cite{Abrams:2012er} and COMET \cite{Kuno:2013mha} have estimated sensitivities of order $10^{-16}$, which can considerably improve the constraints on $|\theta_e\theta_\mu|$. The sensitivity of PRISM/PRIME \cite{Barlow:2011zza} and a Mu2e upgrade \cite{Knoepfel:2013ouy} may even reach $2 \times 10^{-18}$. We can translate this into a sensitivity up to $|\theta_e\theta_\mu| < 3.6 \times10^{-7}/\sqrt{1-0.3 F(x_M)}$.

Furthermore, the sensitivity to the branching ratio for the lepton flavour violating rare tau decay $\tau \to e \gamma$ is expected to improve to $10^{-9}$ at SuperKEKB \cite{Akeroyd:2004mj}, which would improve the sensitivity to the product $|\theta_e\theta_\tau| $ to $|\theta_e\theta_\tau| <  1.5 \times10^{-3}$. 
The role of searches for charged lepton flavour violation in future global fits will depend crucially on whether deviations from the SM are found or not, for instance on whether the best fit value for $| \theta_e |$ will remain non-zero. 
We note that charged lepton flavour violation can also be tested at the $Z$ pole at future lepton colliders, which has recently been studied for the case of the FCC-ee in ref.~\cite{Abada:2014cca}.
%In our discussion of future sensitivities we will therefore focus on the improvements from ``indirect'' and ``direct'' collider searches only. 

Finally, we would like to remark that the processes we considered in this study are only sensitive to the moduli $| \theta_\alpha |$ of the active-sterile mixing angles (or equivalently to the moduli of the Yukawa couplings $| y_{\nu_\alpha} |$). Sensitivity to the phases of the parameters could be achieved in neutrino oscillation experiments, as discussed in the effective theory framework MUV in \cite{FernandezMartinez:2007ms,Antusch:2009pm}.

\section{Discussion and Conclusions}

\begin{figure}
\begin{minipage}{0.49\textwidth}
\begin{center}
\includegraphics[scale=0.55]{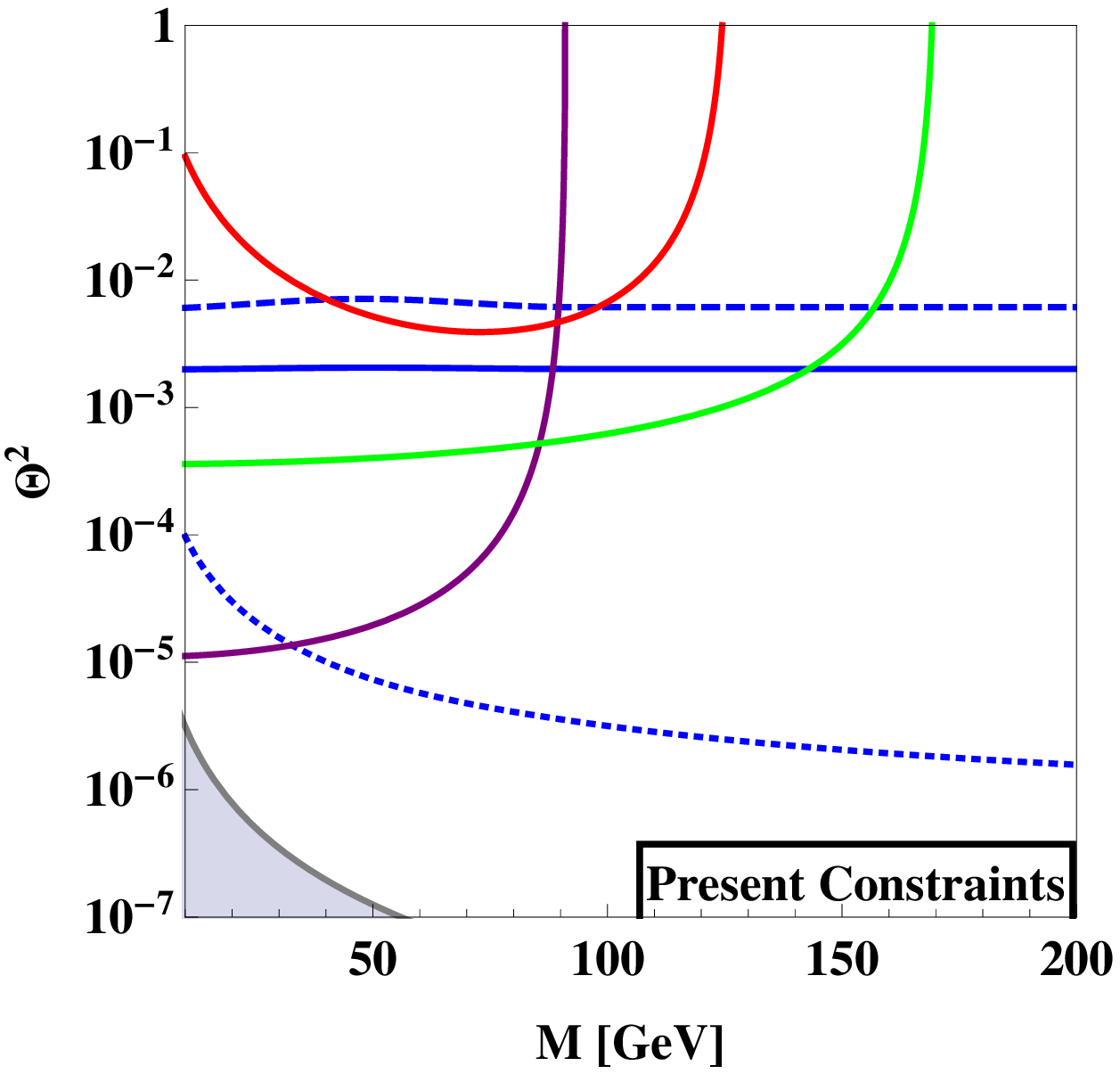}
\end{center}
\end{minipage}
\begin{minipage}{0.49\textwidth}
\begin{center}
\includegraphics[scale=0.55]{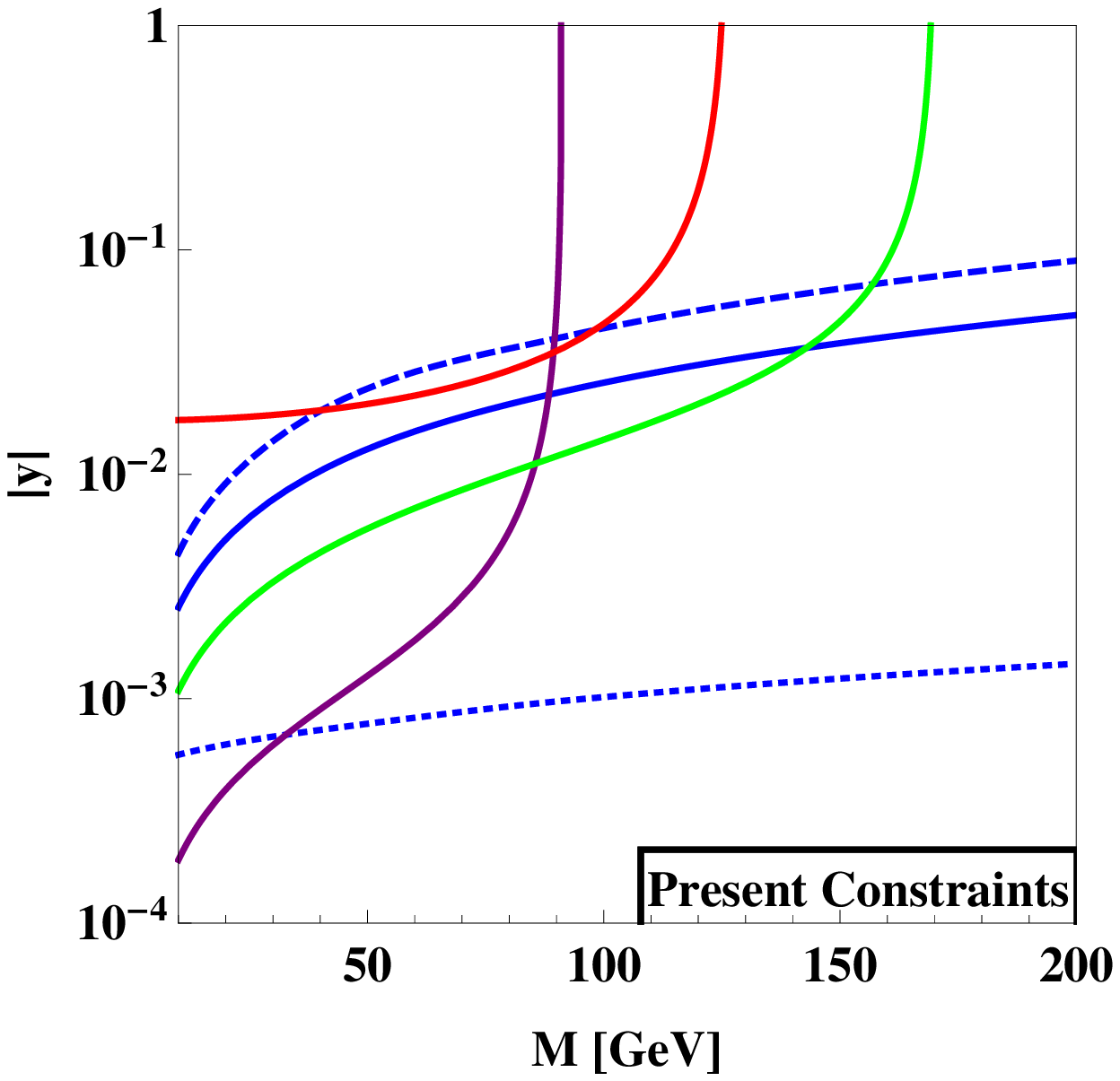}
\end{center}
\end{minipage}

\begin{minipage}{0.49\textwidth}
\begin{center}
\includegraphics[scale=0.8]{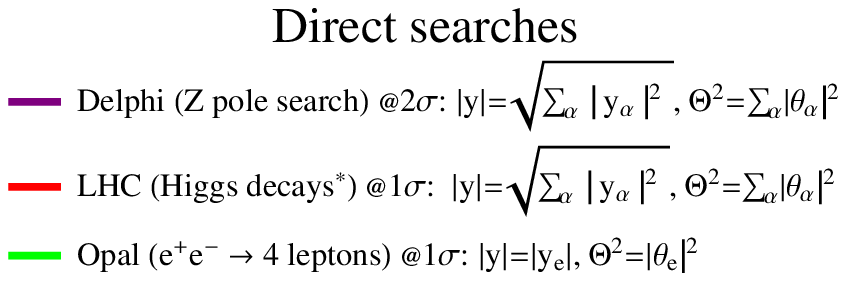}
\end{center}
\end{minipage}
\begin{minipage}{0.49\textwidth}
\begin{center}
\includegraphics[scale=0.8]{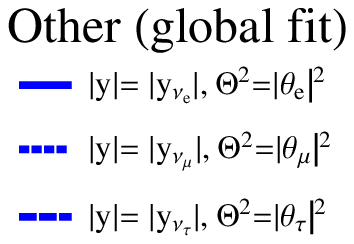}
\end{center}
\end{minipage}

\caption{
Summary of present constraints on sterile neutrino properties. The LHC constraint comes from $h\to\gamma\gamma$.
}
\label{fig:summarypresent}
\end{figure}

\begin{figure}
\begin{minipage}{0.49\textwidth}
\begin{center}
\includegraphics[scale=0.5]{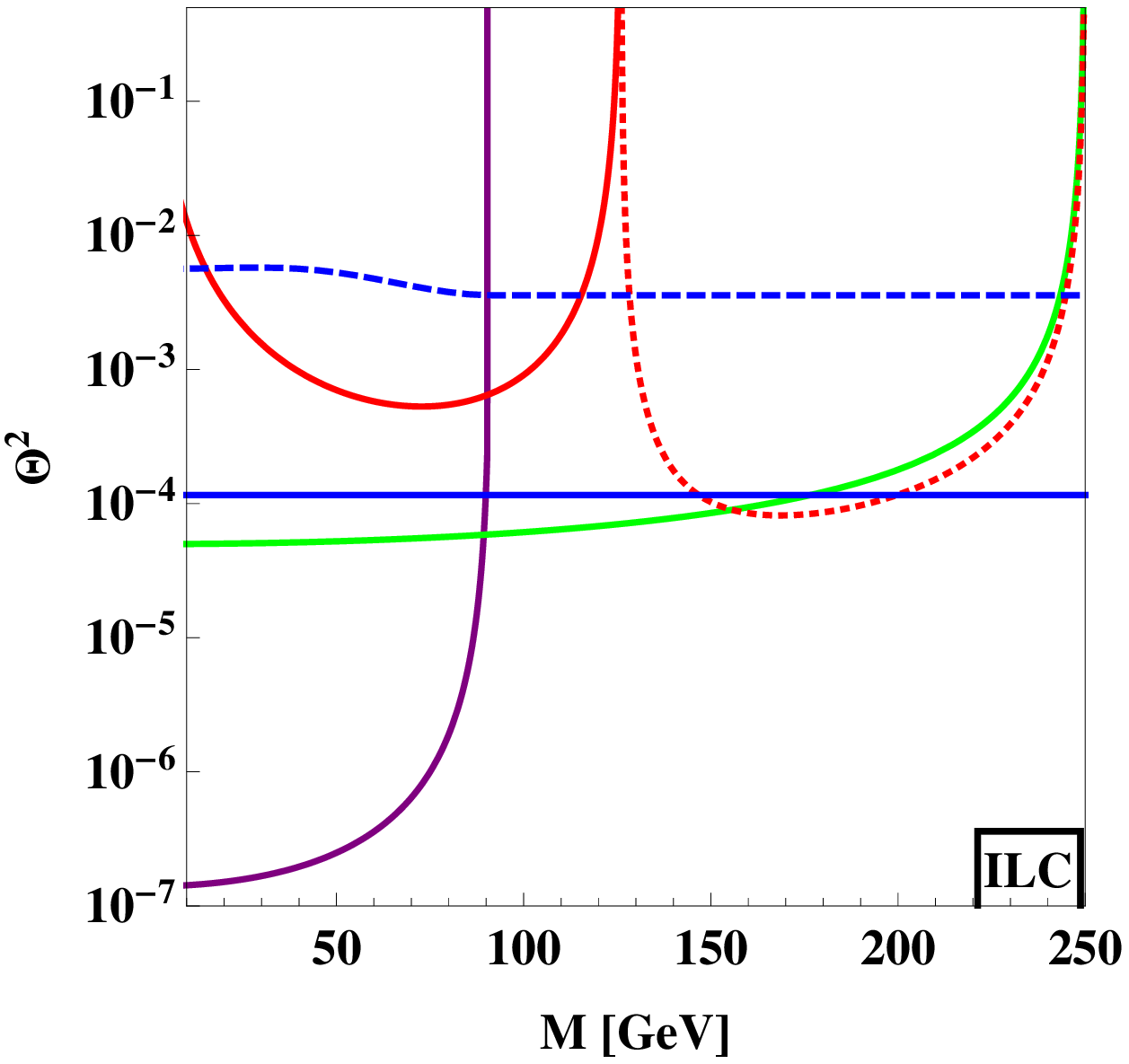}
\end{center}
\end{minipage}
\begin{minipage}{0.49\textwidth}
\begin{center}
\includegraphics[scale=0.5]{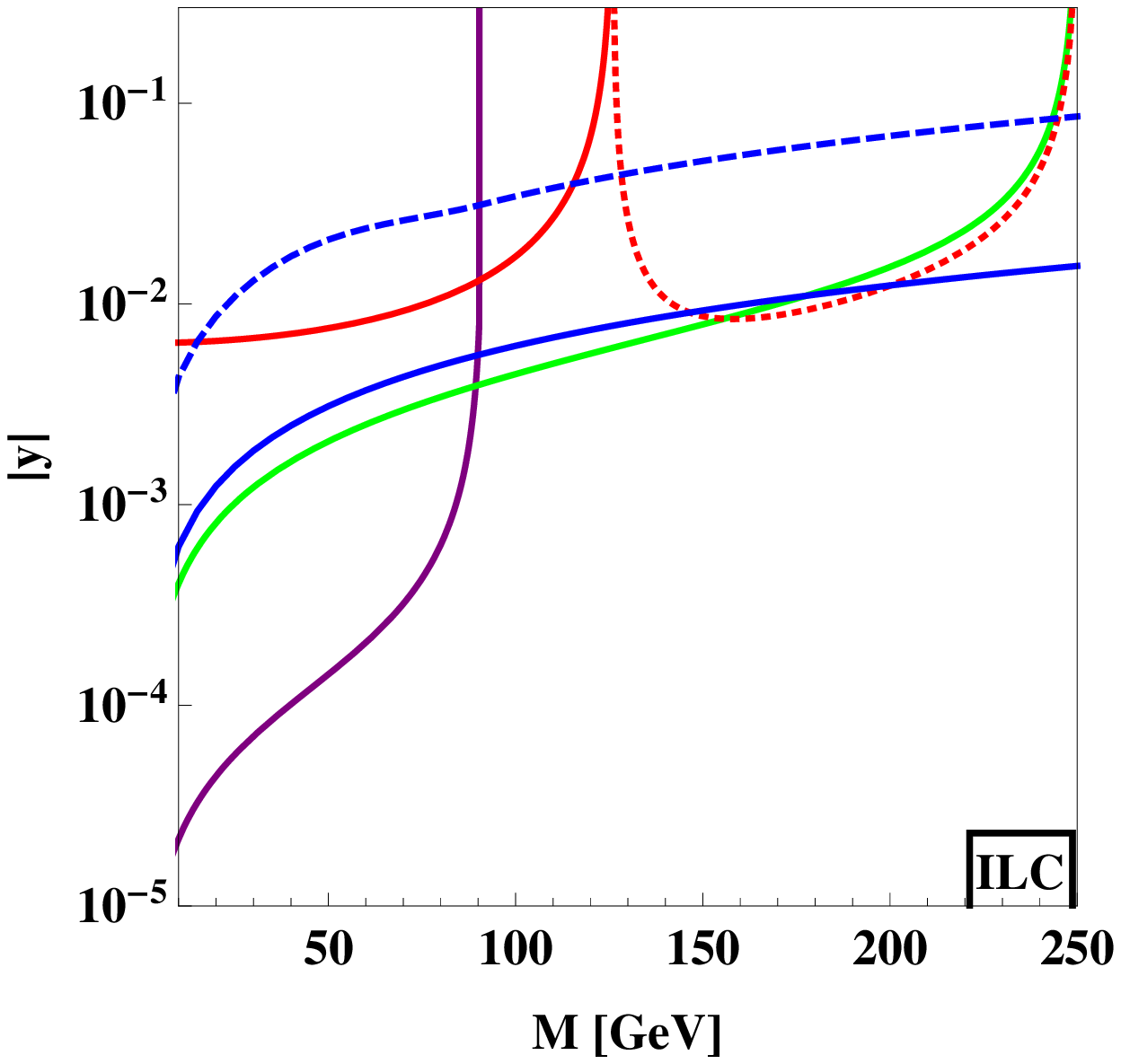}
\end{center}
\end{minipage}

\begin{minipage}{0.49\textwidth}
\begin{center}
\includegraphics[scale=0.5]{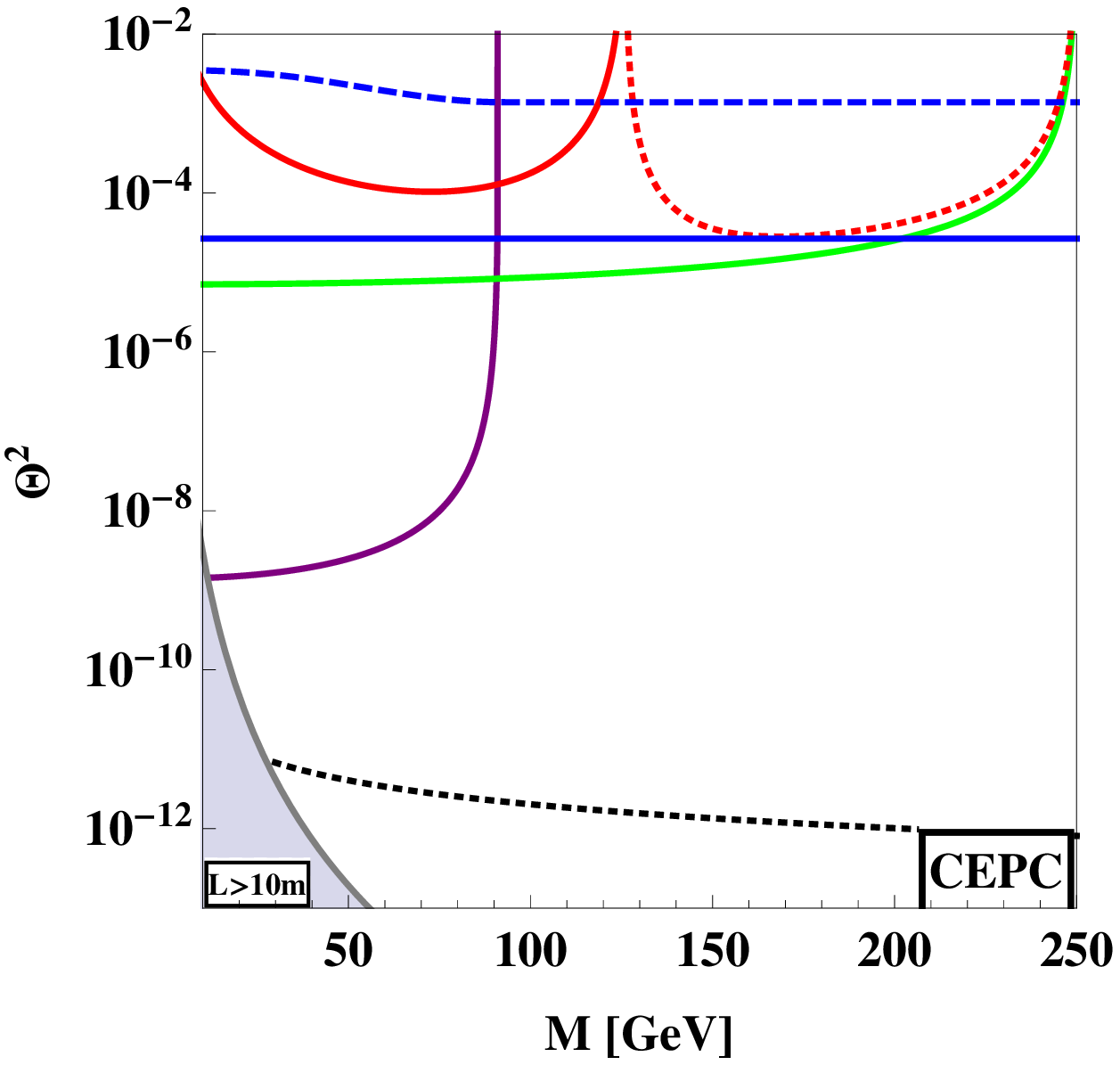}
\end{center}
\end{minipage}
\begin{minipage}{0.49\textwidth}
\begin{center}
\includegraphics[scale=0.5]{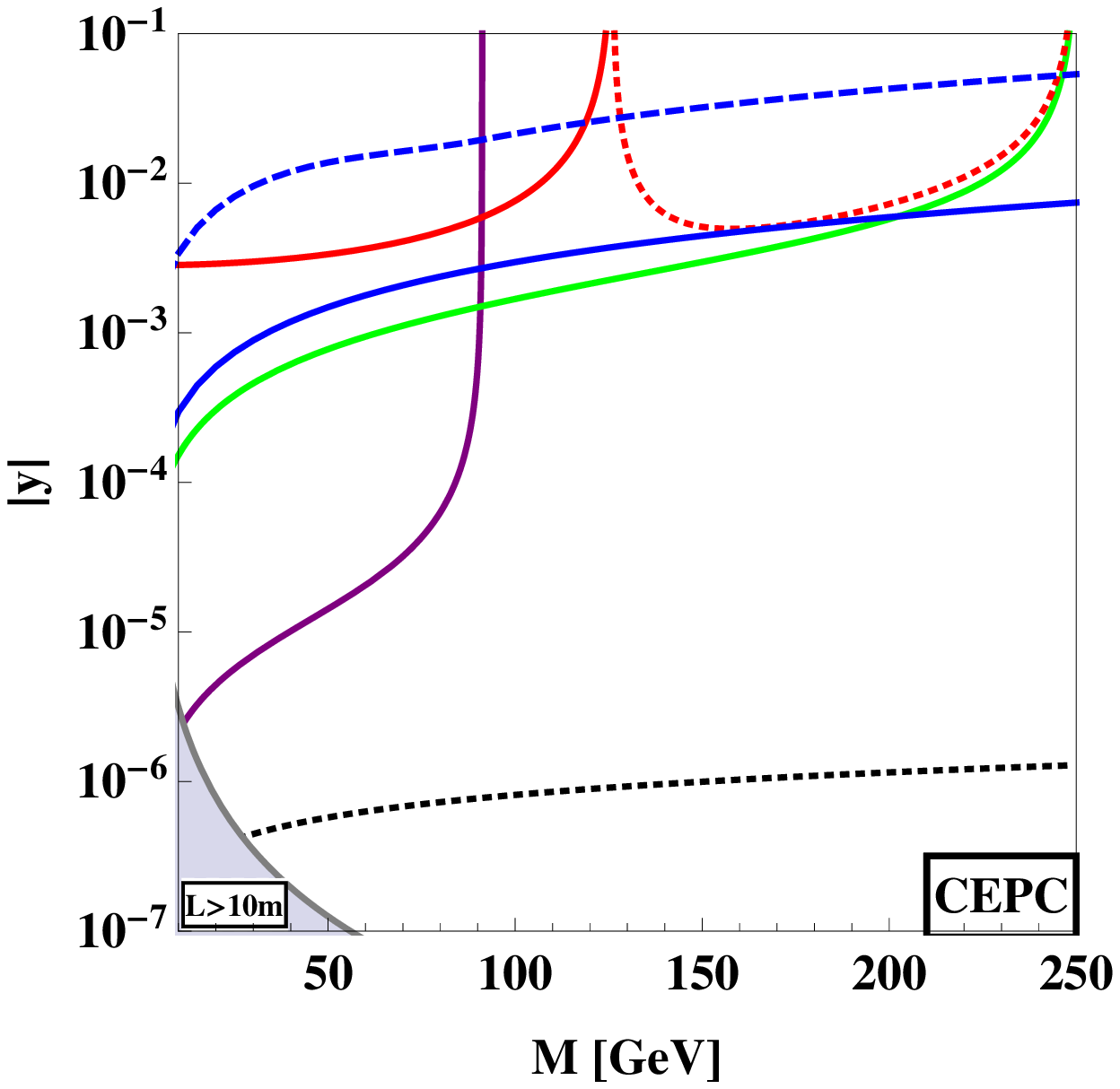}
\end{center}
\end{minipage}

\begin{minipage}{0.49\textwidth}
\begin{center}
\includegraphics[scale=0.5]{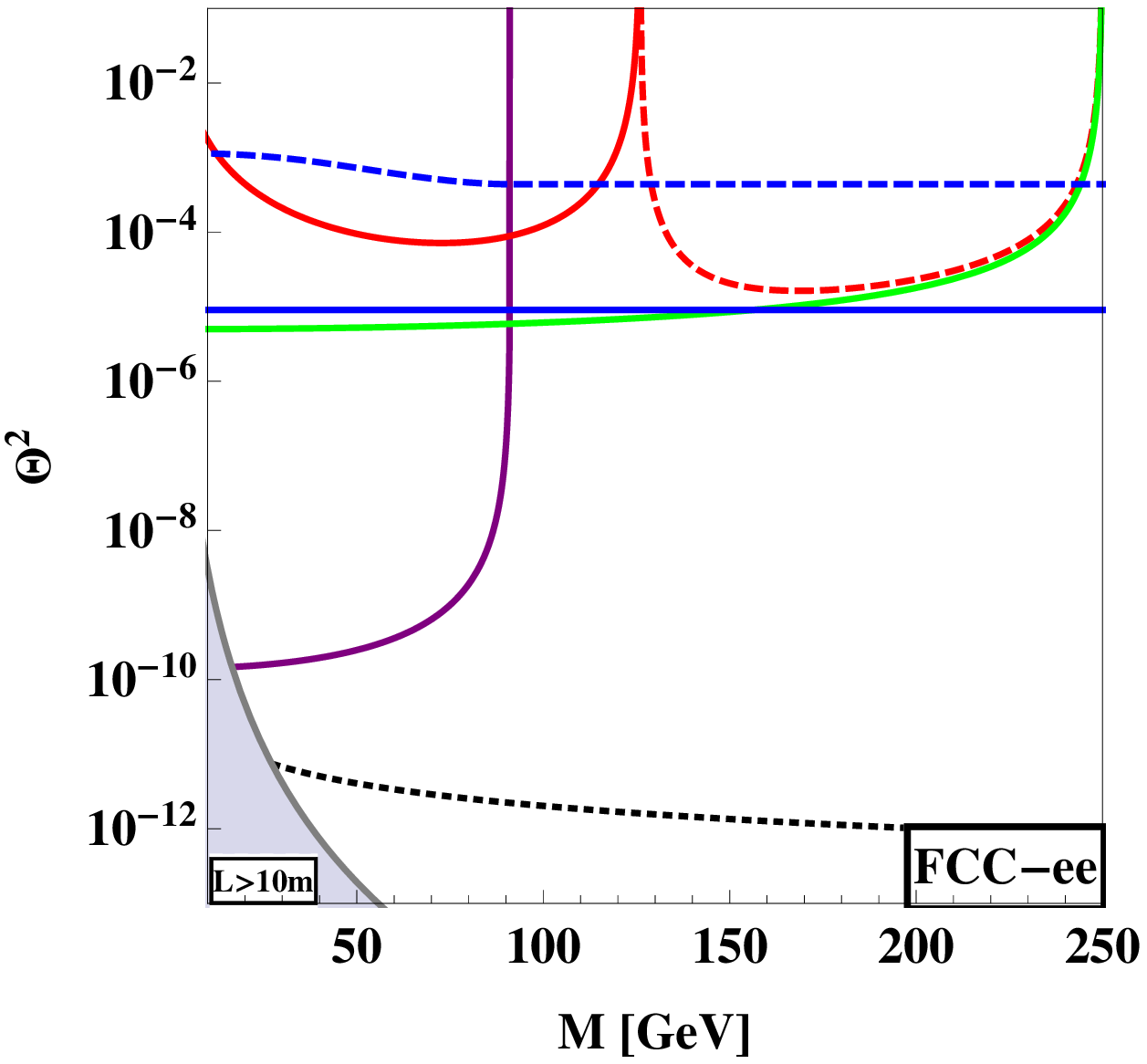}
\end{center}
\end{minipage}
\begin{minipage}{0.49\textwidth}
\begin{center}
\includegraphics[scale=0.5]{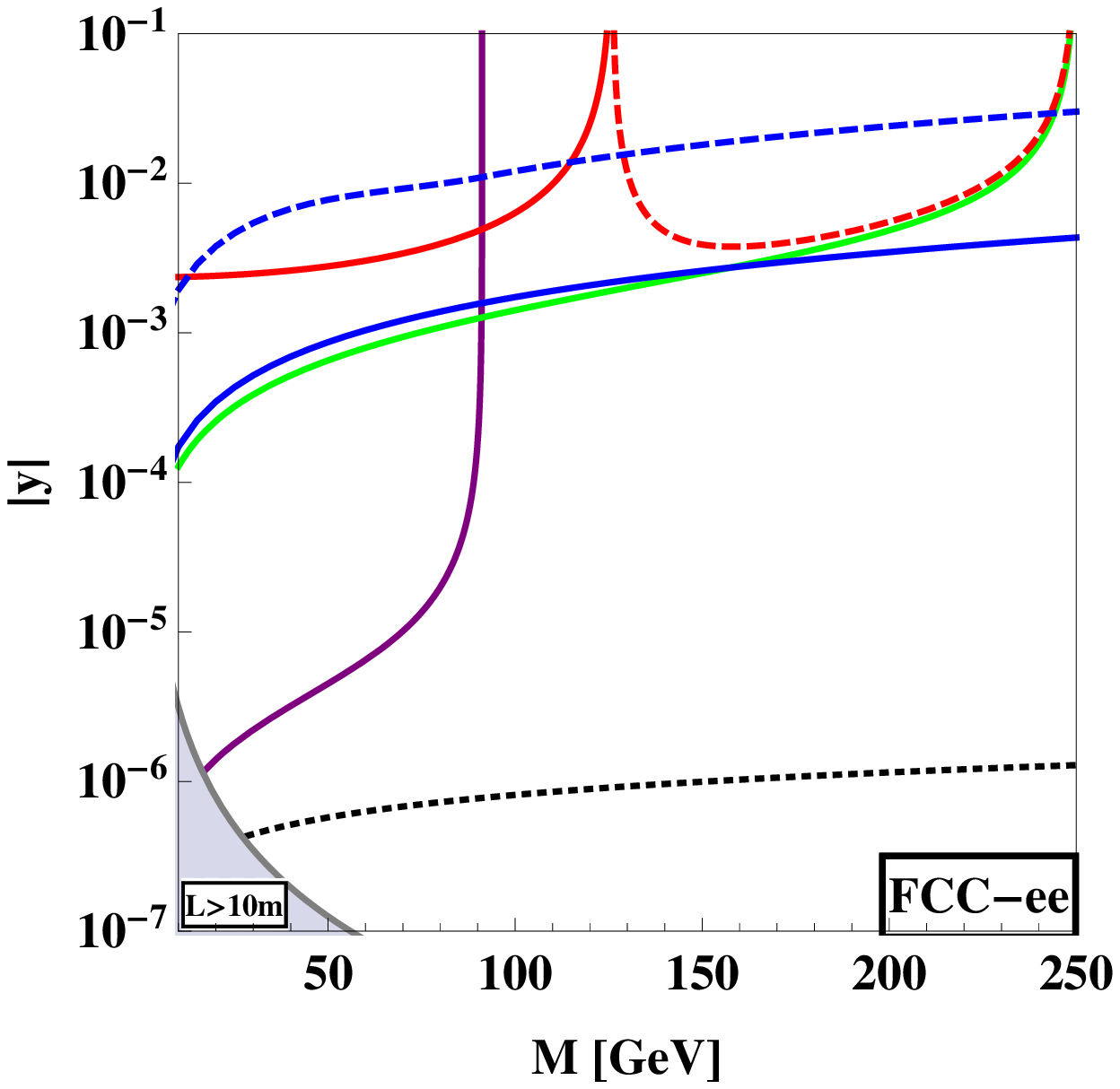}
\end{center}
\end{minipage}

\begin{minipage}{0.49\textwidth}
\begin{center}
\includegraphics[scale=0.8]{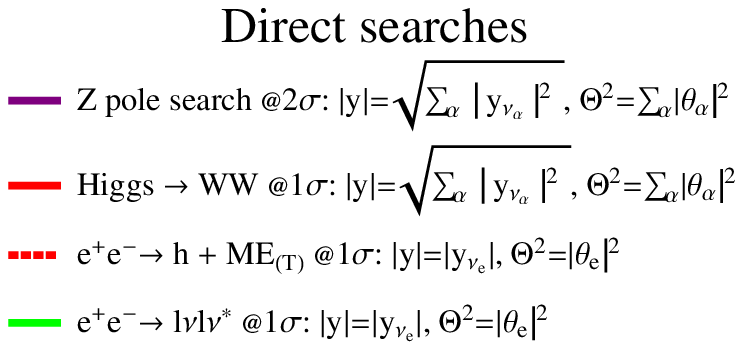}
\end{center}
\end{minipage}
\begin{minipage}{0.49\textwidth}
\begin{center}
\includegraphics[scale=0.8]{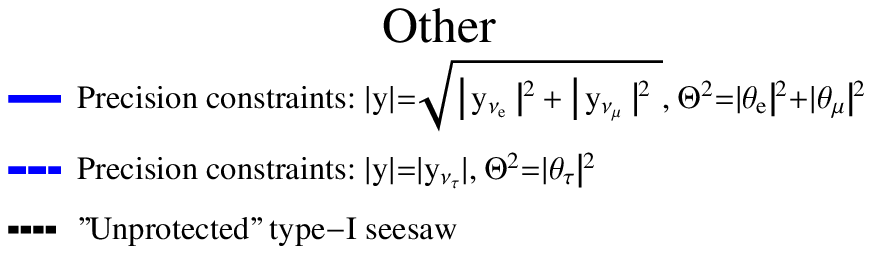}
\end{center}
\end{minipage}

\caption{
Summary of estimated sensitivities to sterile neutrino properties at future colliders. }
\label{fig:colliders}
\end{figure}

We summarize the present constraints and possible future sensitivities on sterile neutrino parameters in the minimal symmetry protected type-I seesaw scenario (cf.\ section 2)
in figs.~\ref{fig:summarypresent} and \ref{fig:colliders}. The scenario allows for a natural realization of sterile neutrino masses $M$ around the EW scale with large, even {\cal O}(1) Yukawa couplings. It is minimal but also sufficiently general so that a distinction between the Yukawa couplings $y_{\nu_\alpha}$  is possible. 
The sterile neutrino parameters of the scenario, to be tested experimentally, are thus $M$ and $y_{\nu_\alpha}$, or equivalently $M$ and the active-sterile mixing angles $\theta_\alpha$ (with the mappings between the parameters summarised in table \ref{tab:pars}).
We like to argue that it is an interesting benchmark scenario for evaluating the present and also the future experimental sensitivities to sterile neutrino properties.

Regarding the present constraints summarized in fig.~\ref{fig:summarypresent}, we note that due to the modification of the Fermi constant measured from $\mu$ decays at low energies, the EWPOs also provide strong constraints for smaller $M$. In agreement with \cite{Antusch:2014woa}, we find a non-zero best fit value for $|\theta_e|$ at 90\% Bayesian confidence level. Nevertheless, to be conservative, we rather present our results only as constraints here. Fig.~\ref{fig:summarypresent} also shows that the measurements of the Higgs branching ratios at the LHC are sensitive to decays into sterile neutrinos (only) in a small range around $m_Z$.

The estimated future sensitivities are shown in fig.~\ref{fig:colliders} for the ILC, CEPC and for the FCC-ee (TLEP). The sensitivities are qualitatively similar, however considering the current proposals the FCC-ee ist the most sensitive. It is interesting to note that a strong Higgs program automatically leads to increased sensitivity for heavy neutrino searches around the $WW$ threshold as a byproduct. The search methods shown in the figure are sensitive to different (combinations of) sterile neutrino parameters. 

Provided that heavy neutrinos with mass in the considered  range and sufficiently large Yukawa couplings exist, one would expect to obtain signals in various processes, which could then be used to discriminate between the active-sterile mixings $| \theta_\alpha |$ and measure/constrain the heavy neutrinos' mass. Furthermore, in addition to the processes considered here, other searches such as for instance the search for displaced vertices as recently studied in \cite{Blondel:2014bra}, can provide additional complementary information. We emphasize that the future sensitivities for direct searches presented here are first estimates only and a careful evaluation of the expected future systematic uncertainties is required for more robust forecasts. 

We also like to note that the possible sensitivity of direct searches at the $Z$ pole at the FCC-ee, and, to a lesser extent also at the CEPC, are closing in on the Yukawa couplings from type-I seesaw models ``without protective symmetry'' where the expectation for the Yukawa couplings follows from the relation $m_\nu \sim \frac{v_{EW}^2 y^2}{2\,M}$ with $m_\nu \lesssim 0.2$ eV. 
The Yukawa couplings and the active-sterile mixing parameters, respectively, are included as dashed black lines in the panels of fig.~\ref{fig:colliders} for comparison.
Finally, for $M$ much larger than the EW scale, only the indirect constraints remain and the sensitivities become independent of $M$ and can alternatively be studied in the effective theory framework as done recently in \cite{Antusch:2014woa}.

\subsection*{Acknowledgements}
This work was supported by the Swiss National Science Foundation. We thank Manqi Ruan and Li Gang for support with the machine performance of the Circular Electron Positron Collider. Furthermore, we thank Alain Blondel for help with event selection filters at LEP and Mike Koratzinos, Serguey Petcov and Roberto Tenchini for valuable discussions.

\bibliographystyle{unsrt}

\end{document}